\documentclass[]{pasj02} 
\usepackage[switch,mathlines]{lineno} 
\usepackage{natbib,graphicx,amsmath,multirow,xspace}
\usepackage{appendix}
\usepackage{xcolor}

\newcommand{\msun}{$M_{\odot}$\xspace}

\newcommand{\fluxcgs}{ergs~s$^{-1}$~cm$^{-2}$\xspace}

\newcommand{\rin}{$R_{\rm in}$\xspace}
\newcommand{\rg}{$R_{g}$\xspace}
\newcommand{\risco}{$R_{\mathrm{ISCO}}$\xspace}
\newcommand{\cps}{counts s$^{-1}$\xspace}
\newcommand{\nustar}{\textit{NuSTAR}\xspace}

\newcommand{\chandra}{\textit{Chandra}\xspace}
\newcommand{\nicer}{\textit{NICER}\xspace}

\newcommand{\xrism}{\textit{XRISM}\xspace}
\newcommand{\source}{GX~340+0\xspace}
\newcommand{\relxillns}{{\sc relxillns}\xspace}

\jyear{2025}
\Received{2025/03/25}
\Accepted{2025/07/08}

\begin{document} 

\title{The Structure of the Relativistic Fe Line in \source as Viewed with \xrism/Resolve, \nicer, and \nustar}

\author{
R. M. \textsc{Ludlam},\altaffilmark{1}\altemailmark\orcid{0000-0002-8961-939X} \email{renee.ludlam@wayne.edu} 
R. \textsc{Ballhausen},\altaffilmark{2,3}\orcid{0000-0002-1118-8470}
P. \textsc{Chakraborty},\altaffilmark{4}\orcid{0000-0002-4469-2518}
E. \textsc{Costantini},\altaffilmark{5}\orcid{0000-0001-8470-749X}
L. \textsc{Corrales},\altaffilmark{6}\orcid{0000-0002-5466-3817}
H. \textsc{Hall},\altaffilmark{1}\orcid{0009-0000-4409-7914}
C. \textsc{Kilbourne},\altaffilmark{3}\orcid{0000-0001-9464-4103}
D. L. \textsc{Moutard},\altaffilmark{6}\orcid{0000-0003-1463-8702}
T. \textsc{Nakagawa},\altaffilmark{7,8}\orcid{0000-0002-6660-9375}
F. S. \textsc{Porter},\altaffilmark{3}\orcid{0000-0002-6374-1119}
I. \textsc{Psaradaki},\altaffilmark{9}\orcid{0000-0002-1049-3182}
M. \textsc{Sudha},\altaffilmark{1}\orcid{0000-0003-0440-7978}
R. K. \textsc{Smith},\altaffilmark{4}\orcid{0000-0003-4284-4167}
H. \textsc{Takahashi},\altaffilmark{10}\orcid{0000-0001-6314-5897}
C. \textsc{Done},\altaffilmark{11}\orcid{0000-0002-1065-7239}
J. A. \textsc{Garc\'{i}a},\altaffilmark{12,13}\orcid{0000-0003-3828-2448}
}
\altaffiltext{1}{Department of Physics \& Astronomy, Wayne State University, 666 West Hancock Street, Detroit, MI 48201, USA}
\altaffiltext{2}{Department of Astronomy, University of Maryland, College Park, MD 20742, USA}
\altaffiltext{3}{NASA / Goddard Space Flight Center, Greenbelt, MD 20771, USA}
\altaffiltext{4}{Center for Astrophysics|Harvard \& Smithsonian, Cambridge, MA, USA}
\altaffiltext{5}{SRON Netherlands Institute for Space Research, Leiden, The Netherlands}
\altaffiltext{6}{Department of Astronomy, University of Michigan, 1085 S. University, MI 48109, USA}
\altaffiltext{7}{Institute of Space and Astronautical Science (ISAS), Japan Aerospace Exploration Agency (JAXA),
3-1-1 Yoshinodai, Chuo-ku, Sagamihara, 252-5210 Kanagawa, Japan}
\altaffiltext{8}{Advanced Research Laboratories, Tokyo City University, 1-28-1 Tamazutsumi, Setagaya-ku, Tokyo 158-8857, Japan}
\altaffiltext{9}{MIT Kavli Institute for Astrophysics and Space Research, Massachusetts Institute of Technology, Cambridge, MA 02139, USA}
\altaffiltext{10}{Department of Physics, Hiroshima University, Hiroshima 739-8526, Japan}
\altaffiltext{11}{Centre for Extragalactic Astronomy, Department of Physics, University of Durham, South Road, Durham DH1 3LE, UK}
\altaffiltext{12}{X-Ray Astrophysics Laboratory, NASA Goddard Space Flight Center, Greenbelt, MD, USA}
\altaffiltext{13}{Cahill Center for Astronomy and Astrophysics, California Institute of Technology, Pasadena, CA 91125, USA}


\maketitle

\begin{abstract}
We present a 152 ks \xrism/Resolve observation of the persistently accreting Z source \source. Simultaneous observations also occurred with \nustar and \nicer for 22.47 ks and 2.7 ks, respectively. The source covered the normal branch to the flaring branching during the observations. The data from all three missions were modeled concurrently for each spectral branch. The superior energy resolution of \xrism/Resolve reveals structure within the iron emission line complex regardless of spectral state. We model the reprocessed Fe K line with a reflection model tailored for thermal illumination of the accretion disk by a neutron star.  
The currently available model encompasses the broad components, but narrow emission features remain at the $\sim5$\% level. These remaining features may be described by the presence of an ionized plasma in the system as has been observed in the Z source Cygnus X-2, but subsequent updates to the reflection model code may be able to explain these features. 

\end{abstract}


\section{Introduction}\label{sec:intro}
Low-mass X-ray binaries (LMXBs) consist of a stellar mass companion  (typically $\lesssim1$ \msun) that transfers material onto a compact object via Roche-lobe overflow. 
Bright, persistently accreting neutron stars (NSs) in LMXBs are divided into two categories based upon their X-ray spectral properties and variability: \lq \lq atoll" and \lq \lq Z" type \citep{hasinger89}. The two classes acquire their name from the tracks that the sources trace out in the hardness-intensity diagram (HID) and color-color diagram. 
A difference in mass accretion rate is thought to be the primary driver between the behavioral difference that atoll and Z sources exhibit, because atoll sources tend toward a lower Eddington luminosity in comparison to the near-Eddington Z sources ($\lesssim0.5$ L$_{\mathrm{Edd}}$ and $\sim 0.5-1.0$ L$_{\mathrm{Edd}}$, respectively: \citealt{vanderklis05}). 
Furthermore, observations of transiently accreting NS LMXBs have exhibited properties of both atoll and Z sources, thus supporting the conclusion of observed differences between the classes being driven by mass accretion rate \citep{homan07, lin09, homan10, Ng2023}. 

Z sources in particular trace out three distinct branches that are named the horizontal, normal, and flaring branches (HB, NB, and FB, respectively) on timescales of hours--weeks. The vertices between branches are named the hard apex (HA) at the transition from HB to NB and soft apex (SA) from NB to FB. 
They can be further divided into two subgroups (Sco-like and Cyg-like) based upon how much time they spend in the different branches. Sco-like sources spend little to no time in the HB and extended periods of time in the FB, whereas Cyg-like sources have a strong HB and spend comparatively less time in the FB \citep{kuulkers97}.
While changes in mass accretion are thought to drive the source evolution between branches, it is debated if the mass accretion rate decreases (e.g., \citealt{church06}) or increases (e.g., \citealt{hasinger90}) from the HB to FB. 

The spectra of bright NS LMXBs are typically soft; dominated by thermal emission from the accretion disk and neutron star and/or boundary layer region \citep{popham2001, inogamov99} with Comptonized emission. The source of Comptonized emission has various proposed origins, such as the accretion disk \citep{white88}, boundary layer \citep{mitsuda89}, or corona region \citep{Sunyaev1991}. The geometry of the corona is elusive, but X-ray polarization measurements with {\it IXPE} \citep{ixpe} are starting to reveal the geometry in different systems (see \citealt{ursini24} for NS LMXBs and references therein). Additionally, reprocessed emission (``reflection") occurs when the surrounding accretion disk is externally illuminated by high-energy X-rays originating from close to the NS. This portion of the spectrum is comprised of a reprocessed continuum with a series of atomic features superimposed. These features become broadened due to Doppler, special, and general relativistic effects in this region \citep{Fabian00}. The strength of these effects become more pronounced with proximity to the compact object. Therefore, these reflection features can be used to infer fundamental properties of the compact object, such as NS radius \citep{Bhattacharyya07, cackett08, Ludlam17a, Ludlam22} or magnetic field strength \citep{ibragimov2009, cackett09, papitto09,  king16, Ludlam17c}, as well as the accretion disk itself (e.g., composition, density, etc.). See \cite{ludlam24} for a recent review and additional references.

\begin{table*}[t!]
\caption{\source observation information for each mission and corresponding instrument considered herein and the effective exposure of the data after filtering into respective branches, i.e., normal branch (NB), soft apex (SA), and flaring branch (FB).}
\label{tab:obs}

\begin{center}
\begin{tabular}{llcccc}
\hline

Mission & Instrument & Sequence ID & Obs.\ Start (UT) & State & Exp.\ (ks) \\
\hline
XRISM & Resolve&  300002010 & 2024-08-17 11:33:18 & NB & 60.2 \\
&  & & & SA & 37.1\\
& & & & FB & 14.3\\
NuSTAR & FPMA & 30901012002 & 2024-08-17 23:16:09 & NB & 7.3 \\
& & & & SA & 8.6\\
& & & & FB & 4.1\\
& FPMB & & & NB & 7.5\\
& & & & SA & 8.8\\
& & & & FB & 4.2\\
NICER & XTI & 7108010105 & 2024-08-18 00:04:14 & NB & 0.85 \\
&  & & & SA & 1.12\\
&  & & & FB & 0.15\\
\hline

\end{tabular}
\end{center}
\end{table*}

\source is a bright, persistently accreting, Cyg-like Z source located at a distance of $11\pm3$~kpc \citep{penninx93, fender00} at an inclination angle of $20^{\circ}\leq i \leq41^{\circ}$ \citep{fender00, dai09, cackett10, miller16} with a high column density along the line of sight ($N_{\rm H}\sim4-11\times10^{22}\ \rm cm^{-2}$: \citealt{lavagetto04, iaria06, dai09, cackett10, chattopadhyay24}). Although no Type-I X-ray bursts have been observed, the source exhibits timing properties that indicate a NS accretor (e.g., twin kHz QPOs: \citealt{jonker98}). The source has recently been observed with {\it IXPE}.
The model-independent X-ray polarization angle ($38^{\circ}\pm6^{\circ}$) agrees with the inclination angle of the system  with a decrease in degree of polarization from the HB to NB \citep{bhargava24a,bhargava24b,lamonaca24}. This suggests geometrical changes in the system contributed to variations in the hard X-ray components (such as the Comptonized emission, blackbody from the NS or boundary layer, or reflection) rather than the accretion disk, but the data are insufficient to determine if a single or combination of components are responsible for the observed polarization behavior \citep{bhargava24b}.
The relativistic Fe~K emission line in \source does not show any strong variation along the horizontal branch and is inferred to arise from close to the last stable circular orbit in the accretion disk \citep{dai09, cackett10}. 
The structure of the Fe line is complex with an apparent  absorption line near 6.9 keV superimposed on the emission line component \citep{dai09, cackett10, miller16}.  From analysis of \chandra data when the absorption was present during the HB, \cite{miller16} utilized photoionized absorption models to account for the feature at $6.94\pm0.02$~keV; deducing that this could represent a modest inflow if it arises from H-like Fe {\sc xxvi} or a strong outflow at $v=0.04c$ if due to He-like Fe {\sc xxv}. If indeed the feature is due to a strong outflow, it would be one of the fastest outflows among NSs and stellar-mass black holes \citep{miller16}.

On the other hand, it is possible to explore the inner accretion flow in more detail and another scenario for the apparent narrow absorption feature. More specifically, the recent development \relxillns \citep{Garcia22}, which is a flavor of {\sc relxill} that is tailored for thermal illumination of the accretion disk by the NS surface or a boundary layer between the surface of the NS and inner edge of the accretion disk, may be able to model this feature. 
\relxillns\ self-consistently connects the angle-dependent reflection spectrum (produced with a blackbody illumination in {\sc xillverNS}) with the ray tracing code {\sc relline} \citep{Dauser10}. At each disk radius, the model chooses the appropriate reflection spectrum for each emission angle calculated in a curved space-time. The resulting reflection model thus accurately captures the detailed dependence of the emitted spectrum with the viewing angle \citep{garcia14}. The model has the advantage of a variable disk density component up to the limit of $10^{19}$ cm$^{-3}$, which is more appropriate for the densities expected in accreting NS systems given that the inner disk should exceed $10^{20}$ cm$^{-3}$ \citep{shakura73, Frank02}. When applied to data that has moderate energy resolution (e.g., 
\nicer's energy resolution of 137 eV at 6 keV), the  Fe {\sc xxv} and Fe {\sc xxvi} K-alpha lines can be produced at a similar strength at disk densities near $10^{19}$ cm$^{-3}$, which creates the appearance of absorption in the broadened Fe K emission line (see figure 3 in \citealt{ludlam24}). 

The majority of existing literature on \source use single emission line models like {\sc diskline} \citep{dai09, cackett10, miller16}  or a Gaussian to account for the Fe K line \citep{lavagetto04, ueda05, church06, bhargava24a, bhargava24b}, rather than a model that accounts for the fully reprocessed reflection spectrum. There are only two recent analyses of \source that have used \relxillns with moderate energy resolution data: one while the source was in the HB using {\sc IXPE}, \nicer, and \nustar data \citep{lamonaca24} and another looking at \nustar data throughout the Z-track \citep{Li25}. Here, we present \xrism/Resolve \citep{xrism} observations of \source taken during the performance verification phase of the mission with simultaneous coordinated observations from \nustar \citep{harrison13} and \nicer \citep{Gendreau12}. We explore the structure of the Fe line in the NB to FB and determine if reflection modeling is sufficient to account for the observed line profile or if additional emission mechanisms are required, which could change the current understanding of the geometry in this system.
In \S 2 we present the observations and data reduction, \S 3 the analysis of the joint observations and results, and conclude with a discussion in \S 4.

\section{Observations and Data Reduction}\label{sec:2}
The sequence IDs, observation dates and start times, and exposure time for each branch are given in Table \ref{tab:obs} for each mission and instrument. Additional details regarding the reduction and division of the data are provided in the following subsections. {\sc heasoft} v.6.34 was utilized in all cases.

\begin{figure*}[t!]
 \begin{center}
  \includegraphics[width=\textwidth]{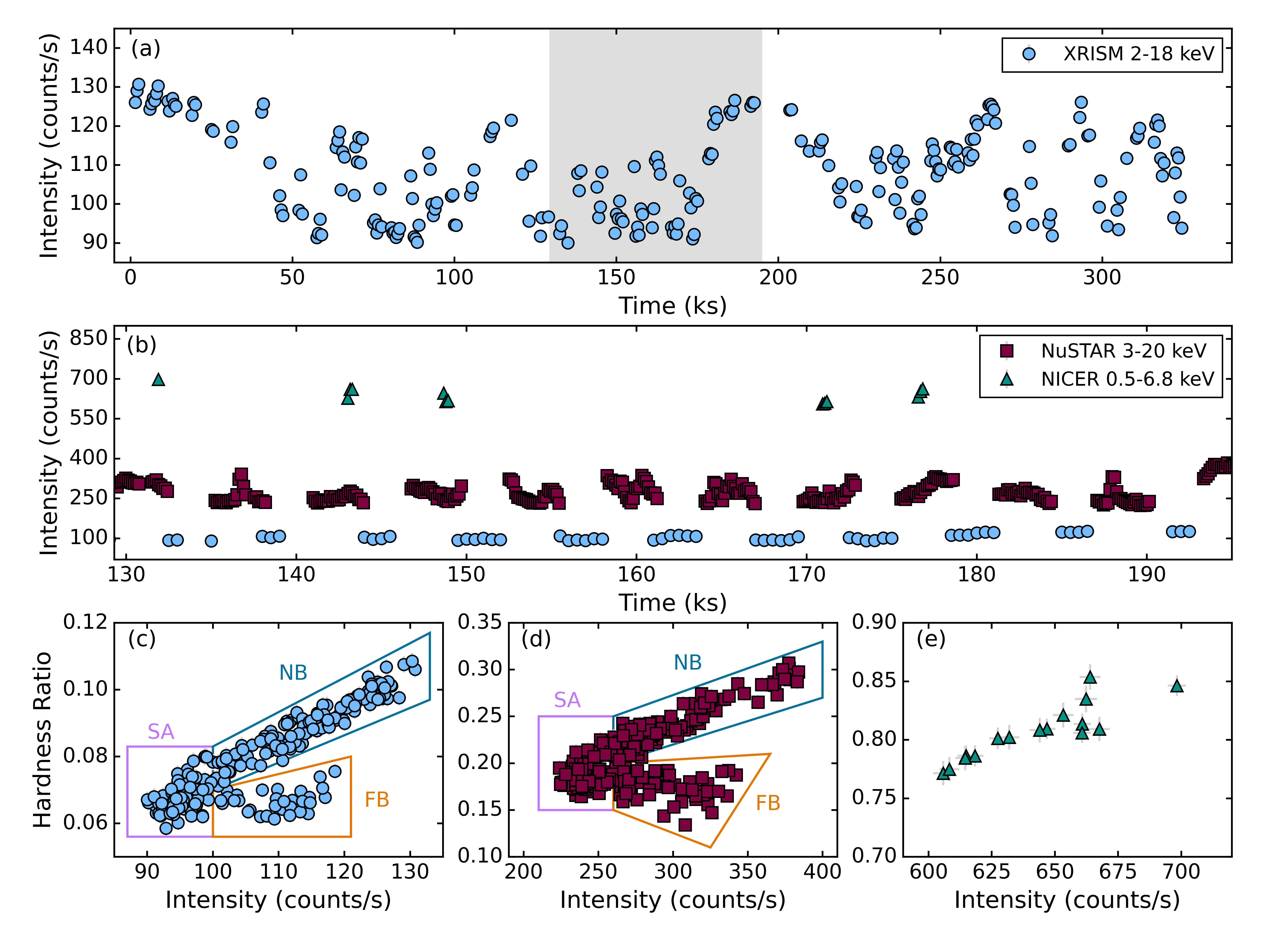} 
 \end{center}
\caption{Light curves and hardness-intensity diagrams for the \xrism (circles), \nustar/FPMA (squares), and \nicer(triangles) observations. Only one FPM is shown for \nustar for clarity. Panel (a) shows the \xrism light curve binned at 500 s in the 2--18 keV energy band. The zero point indicates the start of the \xrism observation (2024-08-17 11:33:18 UT). The shaded gray region indicates where the coordinated observations with \nicer and \nustar occurred. Panel (b) shows the light curve of the \nustar and \nicer data binned to 128 s within the gray region of panel (a). Panels (c), (d), and (e) show the HID of \xrism, \nustar, and \nicer, respectively. The intensity on the x-axis corresponds to the same data shown in panels (a)-(b). The hardness ratio for \xrism and \nustar is constructed from the intensity in the 10--16 keV band to the 6.4--10 keV band. The hardness ratio of the \nicer data is 3.8--6.8 keV band to the 2--3.8 keV band.
{Alt text: Graphs showing the light curve and hardness-intensity diagrams of the \xrism, \nustar, and \nicer data during the observations of \source with subplots labeled a to e.} }\label{fig:hid}
\end{figure*}

\subsection{XRISM}\label{ssec:21}
\xrism observed \source on two occasions (300002010 and 300002020) beginning on August 17th, 2024 and August 31st, 2024 for approximately 152 ks during each epoch. The first observation contains coordinated observations with \nustar and \nicer, thus we only consider this observation hereafter and defer analysis of both observations to a subsequent publication. We follow version 2.1 of the \xrism Quick Start Guide with publicly available CALDB version index 20240815 to reduce the Resolve data. As per the recommendation of the calibration team, pixel 27 was removed from the data and only the highest resolution events (``Hp'') were considered and extracted for light curves and spectra. 
We note that there was small jump ($3.5\pm0.5$~eV at 5.9~keV) in the energy scale of pixel 30 during OBSID 300002010 after the second ADR recycle, affecting 70 ks of elapsed time between calibration measurements. 
Consequently, the energy resolution degrades by $\sim1$ eV for pixel 30 during this time frame. 
This was identified after much of the analysis had been completed through comparison to the gain trend on the other pixels. To investigate the impact of this pixel on the analysis, we compared the data with and without pixel 30 included and found no differences exceeding the statistical errors reported throughout, thus we include the pixel in the subsequent analysis presented. 
Light curves were extracted in the 2--18 keV, 6.4--10 keV, and 10--16 keV energy bands to create the HID (Figure \ref{fig:hid}). GTIs were generated in MET to extract spectra corresponding to the NB, FB, and soft apex (SA) between the two branches\footnote{Separation of the branches into further segments is differed to a follow up analysis.}. Background models for \xrism/Resolve are under development, but given the bright nature of the source, the background is negligible in comparison. 

Given that the count rate incident on the array exceeds 100 counts/s, we investigate the XRISM/Resolve data for known issues impacting bright sources\footnote{See the \xrism Proposal Observatory Guide: Bright Sources for further information.}.
Event loss can occur when the count rate exceeds 200 counts/s and the onboard pulse shape processor (PSP) cannot register all incident events prior to the buffer being reset (\citealt{ishisaki18, mizumoto25, 2025arXiv250606692M}). These periods are logged in the xa300002010rsl\_el.gti.gz file within the \xrism/Resolve unfiltered event folder. We compare the event loss GTIs to the GTIs for each spectral branch to assess what fraction of exposure may be impacted. We find that $\sim$1\% of the time in the NB overlaps with periods flagged for event loss while the FB and SA are unaffected. This can impact the absolute flux measured from the source, however, we account for this with a cross-calibration model during spectral fitting in \S\ref{sec:3}.
Furthermore, we check for cross-talk between electrically neighboring pixels by using the {\sc status==13} expression when selecting events, which removes events likely contaminated by untriggered electrical cross-talk. Direct comparison between the spectra extracted with and without this expression applied provides no noticeable discernible difference in the energy resolution. This is a broad filter that may exclude many events over the array so we also check for cross-talk specifically between the central four pixels (0, 17, 18, 35) and the outer four pixels (1, 16, 19, 34) they are electronically neighboring by extracting spectra from each respective subset and comparing the line features to assess a potential reduction in energy resolution of the Hp events. When directly comparing the Fe line region of the spectrum extracted from the central pixels to the outer pixels, no discernible differences were observed; thus cross-talk does not impact our observation either. 
Since that the gate valve remained closed during the observation, no filter was used to reduce the illuminating flux of the source of the detector. Large response files were generated via {\sc rslmkrmf} and {\sc xaarfgen}.

\subsection{NuSTAR}\label{ssec:22}
\nustar observed \source for 22.47 ks during the first epoch of the \xrism observations. Data were reduced using nustardas v2.1.4 and  CALDB v.20240812. Due to \source being  bright with an excess of 100 \cps, we set `statusexpr=``(STATUS==b0000xxx00xxxx000)\&\&(SHIELD==0)'' in {\sc nupipeline}. 
A region of $100''$ radius centered on the source on DET0 and a background region of the same size but sufficiently far from the source on DET2 were used for spectral and light curve extraction. The light curves were inspected for Type-I X-ray bursts, but none were present. 
The hardness ratio (HR: $10-16$ keV/$6.4-10$ keV) versus $3-20$ keV intensity is shown in Figure \ref{fig:hid}, as well as the light curve relative to the \xrism observation. The \nustar observation traced out the normal to flaring branch in the HID. GTIs were created in MET to extract spectra for the appropriate branches.

\begin{table*}[t!]

\caption{Continuum modeling of the joint \xrism, \nicer, and \nustar data in the normal branch (NB), soft apex (SA), and flaring branch (FB).}
\label{tab:continuum} 

\begin{center}

\begin{tabular}{ll|ccc|ccc}
\hline

Model & Parameter & \multicolumn{3}{c|}{Model 1a} & \multicolumn{3}{c}{Model 2a} \\
& & NB & SA & FB & NB & SA & FB \\
\hline

{\sc crabcor} 
& $C_{\rm FPMB}\ ^{\dagger}$ 
& $ 0.980 \pm 0.001$
& $ 0.980 \pm 0.001$
& $ 0.980 \pm 0.001$
&  $ 0.980 _{- 0.001 }^{+ 0.002 }$
&  $ 0.980 _{- 0.001 }^{+ 0.002 }$
& $ 0.980 _{- 0.001 }^{+ 0.002 }$
\\
& $C_{\rm XRISM}$
& $ 0.696 _{- 0.004 }^{+ 0.005 }$
& $ 0.776 \pm 0.005 $
& $ 0.678 \pm 0.005$
& $ 0.697 \pm 0.004$
&$ 0.776 _{- 0.004 }^{+ 0.005 }$
&$ 0.679 \pm 0.004$

\\
& $C_{\rm NICER}$
& $ 0.749 _{- 0.004 }^{+ 0.002 }$
& $ 0.776 _{- 0.003 }^{+ 0.004 }$
&$ 0.702 _{- 0.004 }^{+ 0.009 }$
& $ 0.747 _{- 0.001 }^{+ 0.004 }$
&$ 0.774 _{- 0.002 }^{+ 0.004 }$
&$ 0.703 _{- 0.006 }^{+ 0.008 }$
\\
& $\Delta \Gamma_{\rm XRISM}\ ^{\dagger}$ ($10^{-3}$)
&$ 4 _{- 3 }^{+ 4 }$
&$ 4 _{- 3 }^{+ 4}$
&$ 4 _{- 3 }^{+ 4 }$
& $ 5 \pm3$
& $ 5 \pm3$
& $ 5 \pm3$
\\

{\sc tbabs} 
&$N_{\mathrm{H}}\ ^{\dagger}$ ($10^{22}$ cm$^{-2}$) 
& $ 8.47 _{- 0.04 }^{+ 0.05 }$
& $ 8.47 _{- 0.04 }^{+ 0.05 }$
& $ 8.47 _{- 0.04 }^{+ 0.05 }$
&$ 8.47 \pm0.05$
&$ 8.47 \pm0.05$
&$ 8.47 \pm0.05$
\\

{\sc thcomp} 

& $kT_{e}$ (keV)
&  $ 2.93 _{- 0.03 }^{+ 0.05 }$
&  $ 3.01 _{- 0.09 }^{+ 0.07 }$
&  $ 123 _{- 80 }^{+ 27 }$
&$ 2.94 _{- 0.03 }^{+ 0.05 }$
&$ 3.07 _{- 0.17 }^{+ 0.06 }$
&$ 57 _{- 48 }^{+ 66 }$

\\
& cov\_frac
&  $ 0.29 _{- 0.03 }^{+ 0.02 }$
&  $ 0.12 _{- 0.02 }^{+ 0.03 }$
&  $ 0.005 _{- 0.003 }^{+ 0.001 }$
&$ 0.547 _{- 0.06 }^{+ 0.13 }$
&$ 0.13 _{- 0.01 }^{+ 0.02 }$
&$ 0.0126 _{- 0.008 }^{+ 0.011 }$
\\

{\sc diskbb} 
& $kT_{\rm in}$ (keV) 
&$ 1.43 _{- 0.06 }^{+ 0.05 }$
&$ 1.12 _{- 0.06 }^{+ 0.02 }$
& $ 0.96 _{- 0.03 }^{+ 0.02 }$
& $ 1.41 _{- 0.04 }^{+ 0.09 }$
& $ 1.13 _{- 0.09 }^{+ 0.04 }$
& $ 0.97 _{- 0.03 }^{+ 0.01 }$
\\

& norm$_{\rm disk}$ 
&$ 170 _{- 15 }^{+ 21 }$
&$ 336 _{- 29 }^{+ 61 }$
&$ 617 _{- 56 }^{+ 68 }$
&$ 166 _{- 23 }^{+ 13 }$
& $ 321 _{- 36 }^{+ 87 }$
& $ 598 _{- 37 }^{+ 85 }$
\\

{\sc bbody} 
& $kT_{\rm bb}$ (keV) 
&$ 1.40 \pm0.03$
&$ 1.39 _{- 0.03 }^{+ 0.02 }$
&$ 1.43 \pm0.01$
&$ 1.42 _{- 0.02 }^{+ 0.04 }$
&$ 1.40 _{- 0.05 }^{+ 0.02}$
&$ 1.43 \pm0.01$
\\
& norm$_{\rm bb}$ ($10^{-2}$)
& $ 3.3 _{- 0.6 }^{+ 0.9 }$
& $ 7.2 _{- 0.2 }^{+ 0.6 }$
& $ 12.7_{- 0.2 }^{+ 0.3 }$
&$ 6.8 _{- 1.5 }^{+ 0.6 }$
&$ 8.3 _{- 0.5 }^{+ 1.0 }$
&$ 12.9 _{- 0.2 }^{+ 0.3 }$
\\

& $F_{\rm unabs,\ 0.5-2\ keV}$ 
&$0.5\pm0.1$
&$0.50_{-0.05}^{+0.10}$
&$0.59_{-0.05}^{+0.07}$
& $0.50_{-0.13}^{+0.06}$
&$0.50_{-0.06}^{+0.15}$
&$0.59_{-0.04}^{+0.09}$
 \\

& $F_{\rm unabs,\ 2-10\ keV}$ 
&$1.4\pm0.3$
&$1.2_{-0.1}^{+0.2}$
&$1.4_{-0.1}^{+0.2}$
&$ 1.4_{-0.4}^{+0.2}$
&$ 1.2_{-0.1}^{+0.4}$
&$ 1.4_{-0.1}^{+0.2}$
 \\
 
& $F_{\rm unabs,\ 10-50\ keV}$ 
&$0.15\pm0.03$
&$0.08_{-0.01}^{+0.02}$
&$0.10\pm0.01$
&$ 0.15_{-0.04}^{+0.02}$
&$ 0.08_{-0.01}^{+0.02}$
&$ 0.10_\pm0.01$
 \\
 
& $F_{\rm unabs,\ 0.5-50\ keV}$ 
&$2.0\pm0.4$
&$1.8_{-0.2}^{+0.4}$
&$2.1\pm0.2$
&$ 2.0_{-0.5}^{+0.2}$
&$ 1.8_{-0.2}^{+0.5}$
&$ 2.1_{-0.1}^{+0.3}$
 \\

\hline
\multicolumn{2}{c}{total fit statistic (dof)} & \multicolumn{3}{c}{  15964.10 (10985)} & \multicolumn{3}{c}{ 15970.15 (10985)} \\ 
\hline

\multicolumn{4}{l}{$^{\dagger}=$ tied between all branches} 

\end{tabular}
\end{center}

\medskip
Note.---  Errors are reported at the 90\% confidence level. Model~1a uses a Comptonized disk component whereas Model~2a assumes the Comptonized photons originate from the single-temperature blackbody. In both cases, the optical depth of the Comptonizing region is fixed at $\tau=10$.
The normalization of {\sc bbody} is defined as $(L/10^{39}\ \mathrm{erg\ s^{-1}})/(D/10\ \mathrm{kpc})^{2}$. 
The normalization of {\sc diskbb} is defined as $(R_{in}/\mathrm{km})^{2}/(D/10\ \mathrm{kpc})^{2}\times\cos{\theta}$. 
The unabsorbed flux, $F_{\rm unabs}$, in various energy bands is given in units of $10^{-8}$ \fluxcgs.

\end{table*}

\subsection{NICER}\label{ssec:23}
\nicer observed the source for 2.7 ks in coordination with \nustar during the first epoch of the \xrism\ performance verification observations. 
Data were reduced with nicerdas 2024-08-18\_V013 and the latest CALDB version index 20240206. The data were cleaned with {\sc nicerl2} allowing the thresh\_range=``-3.0-38'' to allow for data taken during orbit day with the light leak on \nicer. 
Light curves were extracted via {\sc nicerl3-lc} in the standard 0.5--6.8 keV, 3.8--6.8 keV, and 2--3.8 keV energy bands \citep{bult18} to create the HID shown in Figure \ref{fig:hid}. 
The soft bandpass of \nicer does not show a clear pattern in the HID, thus similar to \cite{Ludlam22} data were separated into the appropriate branch by converting the \nustar GTIs to \nicer MET where they overlap in time. The GTIs in \nicer MET were applied using {\sc niextract-events} to divide the event file into respective branches. 
Source spectra and response files were extracted using {\sc nicerl3-spect} specifying the generation of the background spectra files via `Scorpeon' (v23) and 1.5\% systematic errors added to the data from the CALDB.

\section{Analysis and Results}\label{sec:3}
The spectra from \xrism/Resolve, \nustar, and \nicer were modeled simultaneously with {\sc XSPEC} v12.14.1 \citep{arnaud96}. All spectra were optimally binned \citep{kb16} prior to modeling using {\sc ftgrouppha} without specifying a minimum number of counts per bin. There are bins with less than 20 counts within the spectra. Therefore, C-statistics were used for \xrism and \nustar, whereas \nicer data used `pgstat' since the data are poissonian while the background file is Gaussian. Similar to \cite{NGC4151}, we ignore \xrism data below 2.4 keV.  Additionally, we restrict the upper bandpass to 12 keV for which the energy-scale accuracy is $<1$ eV for the full energy range considered\footnote{See the \xrism Proposal Observatory Guide}. The \nustar data are modeled from 3.5--30 keV for the NB and SA and 3.5--24 keV for the FB (after which the X-ray background dominates). The \nicer data are fit from 0.8--10 keV due to the observation occurring during orbit day and the current light leak\footnote{https://heasarc.gsfc.nasa.gov/docs/nicer/analysis\_threads/light-leak-overview/} issue leading to enhanced non-source signal below 1 keV. The absorption column along the line of sight was modeled with {\sc tbabs}, which uses {\sc wilm} abundances \citep{wilms00} and {\sc vern} cross-sections \citep{verner96}. The column density ($N_{\rm H}$) was allowed to vary, but was tied between all spectra regardless of spectral state. Errors are reported at the 90\% confidence level.

To account for the cross calibration difference between missions, we use the `{\sc crabcorr}' multiplicative model (see \citealt{steiner10} for the model originally referred to as `{\sc jscrab}'), which has two parameters: 1.) $\Delta \Gamma$ that multiplies the spectrum by a power-law difference (1/E$^{\Delta \Gamma}$) for variations in spectral slope and 2.) a normalization, $C$, akin to a multiplicative constant for absolute flux differences. The multiplicative constant was fixed at unity for FPMA and free to float for all other spectra. 
We find that the multiplicative constant for \xrism is $\sim 0.7$ with respect to \nustar. 
This factor arises in the `rslmkrmf' tool, which incorporates the Hp fraction, but it is unable to determine the true fraction of events that are Hp because of the presence of a large number of spurious low-grade events. These spurious events cannot be disregarded indiscriminately since the source brightness results in a significant number of real low-grade events (T. Yaqoob on behalf of the \xrism calibration team, private communication). 
$\Delta \Gamma$ is set to 0 for the \nustar spectra and allowed to vary for \xrism. The typical $\Delta \Gamma$ for \nicer with respect to \nustar is $\sim-0.05$ during orbit night, but may vary with the undershoot rate for orbit day data due to the light leak (J. Hare, private communication). The majority of the events in the spectra occurred when the undershoot rate exceeded 300 ct/s, thus when $\Delta \Gamma$ was left as a free parameter for the \nicer spectra the value tends toward \nicer being 20\% harder than \nustar. This is over twice the values reported in literature. Because of this, we fix $\Delta \Gamma=-0.1$ for this \nicer dataset (consistent with the upper bound found in literature for other bright XRBs since the light leak: \citealt{moutard23, hall25}). 
All other parameters are tied between spectra of the same state (i.e., NB, SA, and FB).

\begin{figure}[t!]
 \begin{center}
  \includegraphics[width=0.44\textwidth]{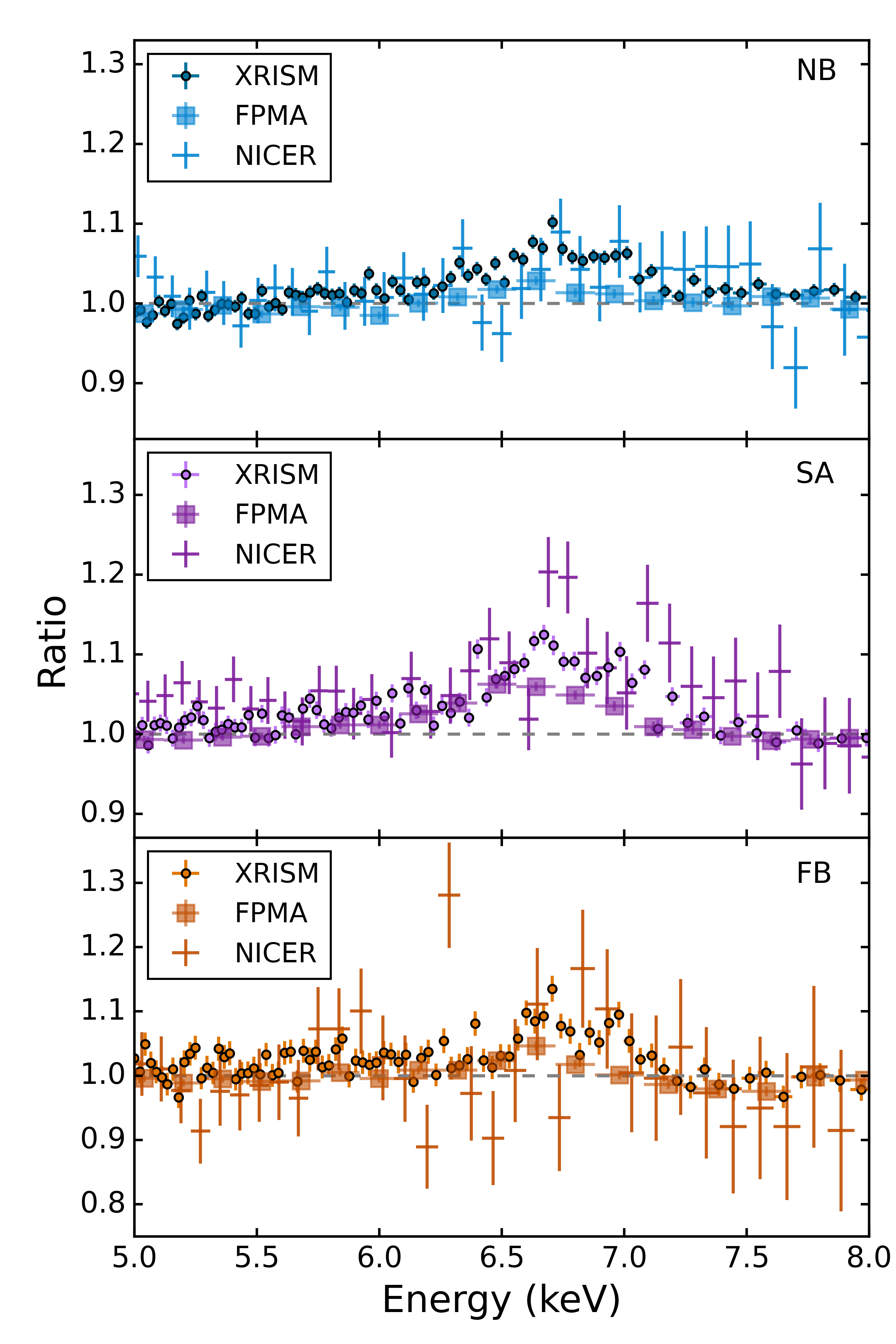} 
 \end{center}
\caption{ Fe line profiles from the ratio of the \xrism/Resolve data in the NB (top), SA (middle), and FB (bottom) to continuum Model 1a. The Fe line profile exhibits a dual-peaked structure. The \nustar/FPMA and \nicer data in each branch are shown for comparison. We note that there is no visual difference in the Fe line profile for Model 1a and Model 2a.
The \xrism data were rebinned for clarity. {Alt text: Graphs showing the iron line profile of the \xrism/Resolve, \nicer, and \nustar data during different branches traced out by \source with subplots labeled a to c.}
}\label{fig:iron}
\end{figure}

To describe the continuum, we start with two thermal components ({\sc diskbb}+{\sc bbody}) to account for emission from the accretion disk and blackbody emission from the NS or boundary/spreading layer region as per \cite{dai09, cackett10}. This provided an apt description of the spectrum up to $15$ keV but the data diverge at $>30\%$ level towards higher energies in the \nustar spectra. 
We account for Comptonization of the high-energy portion of the spectra by applying {\sc thcomp} \citep{thcomp}. Given that this model is a convolution kernal, we extend the energy range for model computation to 0.1 keV to 1000 keV using 1000 logarithmic bins when fitting the data in their respective energy bands as previously stated. We try the convolution on each thermal components as both descriptions have been utilized for \source in recent literature (e.g., \citealt{chattopadhyay24, lamonaca24, bhargava23, bhargava24a, bhargava24b}). We refer to the continuum prescription using the accretion disk for the seed photons (i.e., {\sc thcomp}*{\sc diskbb}) as Model 1a and when the blackbody is the input for the Comptonized photons (i.e., {\sc thcomp}*{\sc bbody}) as Model 2a. When allowing the photon index to be a free parameter when fitting the Comptonized emission within the spectra from the various branches, it would tend to pegged to the lowest model value of $\Gamma=1.001$ in the FB and poorly constrained in the other branches. While a harder photon index in the FB of Z sources is not unprecedented (e.g., $1.3<\Gamma<2$ for the FB of Sco X-1: \citealt{titarchuk14}), this is an uncharacteristically hard value of $\Gamma$. As was done in \cite{chattopadhyay24} for the broadband AstroSat spectral analysis of \source, we set the {\sc thcomp} model to use optical depth instead of $\Gamma$ and fix $\tau=10$ in the subsequent fits. The electron temperature $kT_{e}$ and cover fraction (cov\_frac) are free to vary.
We note that when $\tau$ is left as a free parameter the value is $>1$ (though poorly constrained) indicating an optically thick Comptonizing region as reported in \cite{bhargava23} and \cite{lamonaca24}. We emphasize that the primary focus of this initial study is on the structure of the Fe line region and full investigation of the broadband continuum modeling of the source with different descriptions for the Comptonized emission is outside the scope of this work and deferred to a follow-up publication. Results of applying these continuum models to the spectra in all branches are presented in Table \ref{tab:continuum}.
In both cases, the addition of the thermal Comptonization provides an improvement to the overall fit of the high-energy data above 15 keV (total fit statistic decreased by 419.07 for Model 1a and by 413.02 for Model 2a for 6 degrees of freedom) with comparable fit statistics for both continuum descriptions. Although the fit statistic is comparable between the two models, we note that from the statistical metric of the Bayesian Information Criterion (BIC) for each fit, there is strong evidence ($\Delta \rm BIC=6.05$) that Model 1a is preferred over Model 2a. We continue with both models in an effort to determine how this may impact the modeling of the structure of the Fe line region.

\begin{table}[t]

\caption{Summary of Gaussian line values in the Fe K complex.}
\label{tab:gauss} 

\begin{center}

\begin{tabular}{lccc}
\hline

Gaussian & NB & SA & FB \\
\hline
$E_{1}\ ^{\dagger}$ (keV)
& 6.4 
& 6.4 
& 6.4 
\\
$\sigma_1$ (eV)
&$ 840 _{- 61 }^{+ 57 }$
&$ 411 _{- 77 }^{+ 33}$
&$ 277 _{- 52 }^{+ 81 }$
\\
$K_1$ ($10^{-3}$)
&$ 9.3 _{- 1.2 }^{+ 0.7 }$
&$ 4.2 _{- 1.0 }^{+ 0.5 }$
&$ 1.8_{- 0.2 }^{+ 0.5 }$
\\
$E_{2}\ ^{\dagger}$ (keV)
& 6.7 
& 6.7 
& 6.7 \\
$\sigma_2$ (eV)
&$ 100 _{- 26 }^{+ 21 }$
&$ 143 _{- 17 }^{+ 24 }$
&$ 66 _{- 25 }^{+ 24 }$
\\
$K_2$ ($10^{-3}$)
&$ 1.1 _{- 0.2 }^{+ 0.3 }$
&$ 1.7 _{- 0.2 }^{+ 0.4 }$
&$ 1.3 _{- 0.2 }^{+ 0.4 }$
\\
$E_{3}\ ^{\dagger}$ (keV)
& 6.97 
& 6.97 
& 6.97 \\

$\sigma_3$ (eV)
&$ 28 _{- 8 }^{+ 9 }$
&$ 29 _{- 5}^{+ 13 }$
&$ 0.6 _{- 0.1 }^{+ 0.3 }$

\\
$K_3$ ($10^{-3}$)
&$ 0.20 \pm0.04$
&$ 0.28 _{- 0.03 }^{+ 0.07 }$
&$ 0.4 _{- 0.2 }^{+ 0.1 }$
\\

\hline

\multicolumn{4}{l}{$^{\dagger}=$ fixed} 

\end{tabular}
\end{center}

\medskip
Note.---  Errors are reported at the 90\% confidence level.  The line values are obtained from using Model 1a with three Gaussian lines added. Values from using Model 2a are consistent within the 90\% confidence level, thus we only report the values from one underlying continuum description for clarity. The line width is indicated by $\sigma$. 
The normalization, $K$, of {\sc gauss} is defined as the total photons/cm$^2$/s in the line.

\end{table}

\begin{table*}[t!]

\caption{Reflection modeling of the joint \xrism, \nicer, and \nustar data in the normal branch (NB), soft apex (SA), and flaring branch (FB).}
\label{tab:reflection} 

\begin{center}

\begin{tabular}{ll|ccc|ccc}
\hline


Model & Parameter & \multicolumn{3}{c|}{Model 1b} & \multicolumn{3}{c}{Model 2b}\\
& & NB & SA & FB & NB & SA & FB\\
\hline

{\sc crabcor} 
& $C_{\rm FPMB}\ ^{\dagger}$ 
& $0.981 \pm 0.001$
& $0.981 \pm 0.001$
& $0.981 \pm 0.001$
& $ 0.981 _{- 0.002 }^{+ 0.001 }$
& $ 0.981 _{- 0.002 }^{+ 0.001 }$
& $ 0.981 _{- 0.002 }^{+ 0.001 }$ 
\\
& $C_{\rm XRISM}$
& $ 0.690 \pm 0.001 $
& $ 0.769 _{- 0.002 }^{+ 0.001 }$
& $ 0.674 _{- 0.003 }^{+ 0.001 }$
&$ 0.690 \pm 0.001$
&$ 0.768 _{- 0.001 }^{+ 0.002 }$
&$ 0.672 _{- 0.001 }^{+ 0.003 }$
\\
& $C_{\rm NICER}$
& $ 0.745 _{- 0.003 }^{+ 0.004 }$
& $ 0.772 _{- 0.003 }^{+ 0.004 }$
&  $ 0.699 _{- 0.005 }^{+ 0.008 }$
&$ 0.744 _{- 0.002 }^{+ 0.004 }$
&$ 0.773 _{- 0.004 }^{+ 0.002 }$
&$ 0.701 _{- 0.005 }^{+ 0.010 }$
\\
& $\Delta \Gamma_{\rm XRISM}\ ^{\dagger}$ ($10^{-3}$)
&$ -0.6 _{- 0.2 }^{+ 0.1 }$
&$ -0.6 _{- 0.2 }^{+ 0.1 }$
&$ -0.6 _{- 0.2 }^{+ 0.1 }$
& $ -0.8 _{- 0.1}^{+ 0.3 }$
& $ -0.8 _{- 0.1}^{+ 0.3 }$
& $ -0.8 _{- 0.1}^{+ 0.3 }$
\\

{\sc tbabs} 
&$N_{\mathrm{H}}\ ^{\dagger}$ ($10^{22}$ cm$^{-2}$) 
& $ 8.75 _{- 0.03 }^{+ 0.05}$
& $  8.75 _{- 0.03 }^{+ 0.05}$
& $  8.75 _{- 0.03 }^{+ 0.05}$
& $ 8.72 \pm0.04$
& $ 8.72 \pm0.04$
& $ 8.72 \pm0.04$
\\

{\sc thcomp} 

& $kT_{e}$ (keV)
&  $ 2.86 _{- 0.04 }^{+ 0.05 }$
&  $ 3.27 _{- 0.16 }^{+ 0.08 }$
&  $ 25.1 _{- 10.8 }^{+ 0.4 }$
&  $ 2.86\pm0.06$
&  $ 3.27_{- 0.19 }^{+ 0.09 }$
&  $ 33.4 _{- 1.4 }^{+ 0.6 }$

\\
& cov\_frac 
&  $ 0.28 _{- 0.03 }^{+ 0.02 }$
&  $ 0.06 _{- 0.01 }^{+ 0.02 }$
&  $ 0.0014 _{- 0.0002 }^{+ 0.0001 }$
&  $ 0.73 _{- 0.13 }^{+ 0.07 }$
&  $ 0.10 _{- 0.01 }^{+ 0.03 }$
&  $ 0.007 \pm 0.001 $
\\

{\sc diskbb} 
& $kT_{\rm in}$ (keV) 
& $ 1.42 _{- 0.02 }^{+ 0.03 }$
& $ 1.19 _{- 0.04 }^{+ 0.01 }$
& $ 0.95 _{- 0.03 }^{+ 0.01 }$
& $ 1.43 _{- 0.04 }^{+ 0.04 }$
& $ 1.17 _{- 0.04 }^{+ 0.01 }$
& $ 0.97 _{- 0.03 }^{+ 0.01 }$
\\

& norm$_{\rm disk}$ 
& $ 193 _{- 10 }^{+ 8 }$
& $ 316 _{- 6 }^{+ 39}$
& $ 727 _{- 47 }^{+ 94 }$
& $ 182 _{- 12 }^{+ 13 }$
& $ 329 _{- 14 }^{+ 33}$
& $ 650 _{- 27 }^{+ 93 }$
\\

{\sc bbody} 
& $kT_{\rm bb}$ (keV) 
& $ 1.52 _{- 0.06 }^{+ 0.09 }$
& $ 1.50 _{- 0.04 }^{+ 0.01 }$
&  $ 1.44 \pm0.01$
&$ 1.55 _{- 0.09 }^{+ 0.07 }$
&$ 1.49 _{- 0.04 }^{+ 0.01 }$
&$ 1.45 \pm0.01$
\\
& norm$_{\rm bb}$ ($10^{-2}$)
& $ 1.6\pm0.1$
& $ 5.6 _{- 0.1 }^{+ 0.6 }$
& $ 12.4 \pm0.2$
& $ 5.1 _{- 0.6 }^{+ 0.7 }$
&$ 6.8 _{- 0.2 }^{+ 0.5 }$
&$ 12.1 _{- 0.1 }^{+ 0.3 }$
\\

{\sc relxillNS} 
& $q$
& $ 2.2 _{- 0.2 }^{+ 0.3 }$
& $ 2.1 _{- 0.3 }^{+ 0.2 }$
& $ 2.3 _{- 0.4 }^{+ 0.5 }$
&$ 2.2 \pm0.2$
&$ 2.1 \pm0.2$
&$ 2.4 _{- 0.3 }^{+ 0.7 }$
\\

& $i\ ^{\dagger}$ ($^{\circ}$) 
& $ 39 _{- 5 }^{+ 1 }$ 
&$ 39 _{- 5 }^{+ 1 }$ 
&$ 39 _{- 5 }^{+ 1 }$  
&$ 40 _{- 6}^{+ 1 }$
&$ 40 _{- 6}^{+ 1 }$
&$ 40 _{- 6}^{+ 1 }$
\\

& $R_{\rm in}$ ($R_{\rm ISCO}$)
& $ 1.8 _{- 0.4 }^{+ 0.2 }$
& $ 1.5 _{- 0.4 }^{+ 0.8 }$
& $ 1.6 _{- 0.3 }^{+ 0.7 }$
& $ 1.1 _{- 0.1 }^{+ 0.5 }$
& $ 1.2 _{- 0.2 }^{+ 0.9 }$
& $ 1.2 _{- 0.2 }^{+ 0.5 }$
\\

& $\log (\xi)$ 
& $ 2.3 \pm0.1$
& $ 2.1 \pm0.1$
& $ 1.9 _{- 0.3 }^{+ 0.1 }$
& $ 2.2 _{- 0.1 }^{+ 0.2 }$
& $ 2.0 \pm0.1$
& $ 2.0 _{- 0.2 }^{+ 0.1 }$
\\

& $A_{\rm Fe}\ ^{\dagger}$ 
& $ 8.4 _{- 1.7 }^{+ 0.9 }$ 
& $ 8.4 _{- 1.7 }^{+ 0.9 }$  
& $ 8.4 _{- 1.7 }^{+ 0.9 }$  
& $ 9.7 _{- 0.4 }^{+ 0.3 }$
& $ 9.7 _{- 0.4 }^{+ 0.3 }$
& $ 9.7 _{- 0.4 }^{+ 0.3 }$
\\

& norm$_{\rm refl}$ ($10^{-3}$) 
& $ 1.7 _{- 0.2 }^{+ 0.3 }$
& $ 2.1 _{- 0.3 }^{+ 0.1 }$
& $ 1.9 _{- 0.2 }^{+ 0.4 }$
& $ 1.9 _{- 0.3 }^{+ 0.1 }$
& $ 2.1 _{- 0.3 }^{+ 0.3 }$
& $ 2.16 _{- 0.20 }^{+ 0.03 }$
\\

\hline
\multicolumn{2}{c}{total fit statistic (dof)} & \multicolumn{3}{c}{ 14267.38 (10971)} & \multicolumn{3}{c}{ 14279.60 (10971)}\\ 
\hline

\multicolumn{6}{l}{$^{\dagger}=$ tied between all branches} 

\end{tabular}
\end{center}

\medskip
Note.---  Errors are reported at the 90\% confidence level. Model~1b refers to the continuum model that uses a Comptonized disk component whereas Model~2b assumes the Comptonized photons originate from the single-temperature blackbody. For \relxillns, the disk density is fixed at $\log(n_{e}/\rm cm^{-3})=19$, the dimensionless spin at $a=0$, and reflection fraction at $f_{refl}=-1$ so that the model returns the reflection spectrum only. For $a=0$, the value of $1\ R_{\rm ISCO}=6\ R_{g}$.

\end{table*}

\begin{figure*}[t!]
 \begin{center}
  \includegraphics[width=0.95\textwidth]{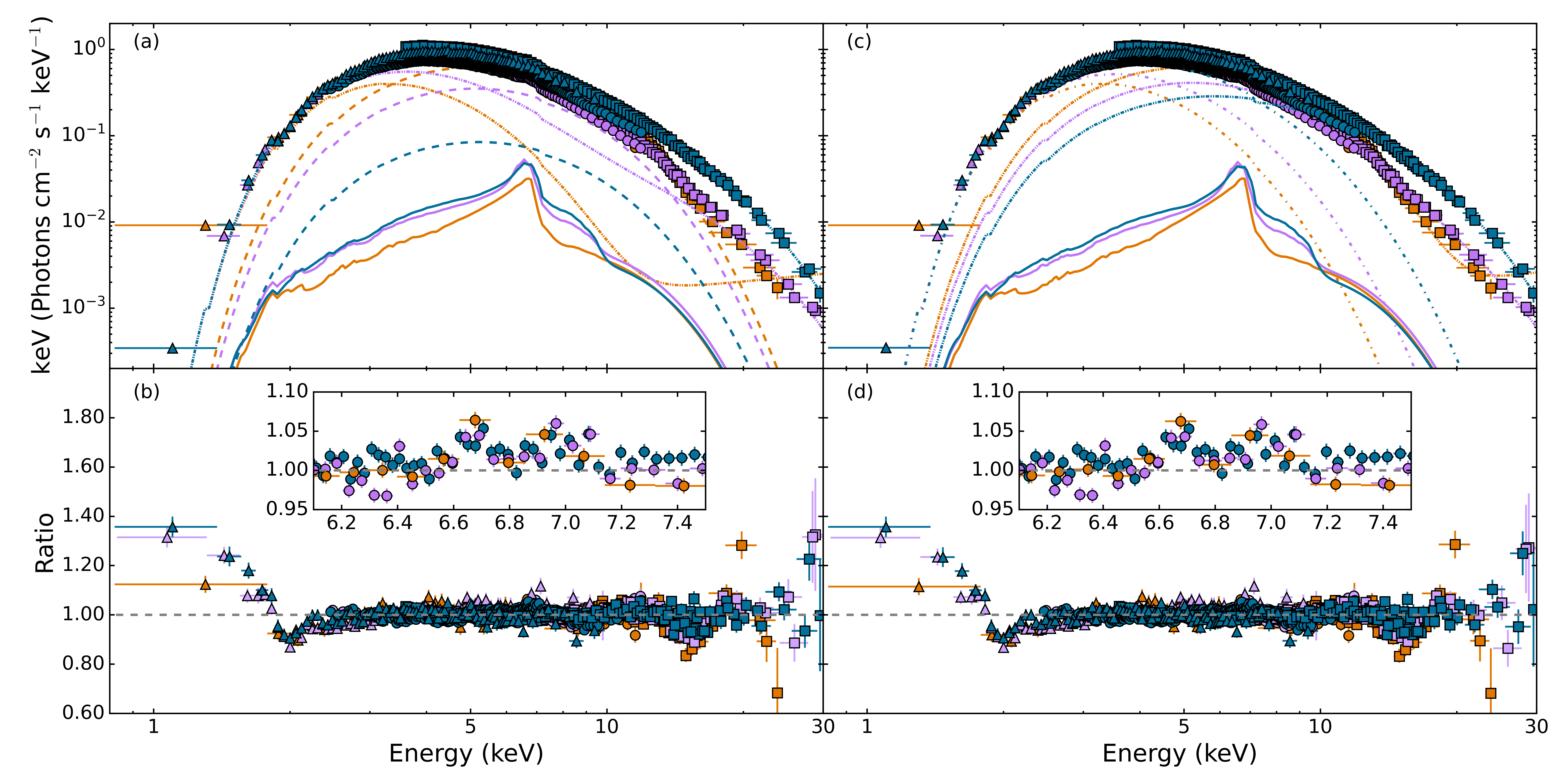} 
  \includegraphics[width=0.95\textwidth]{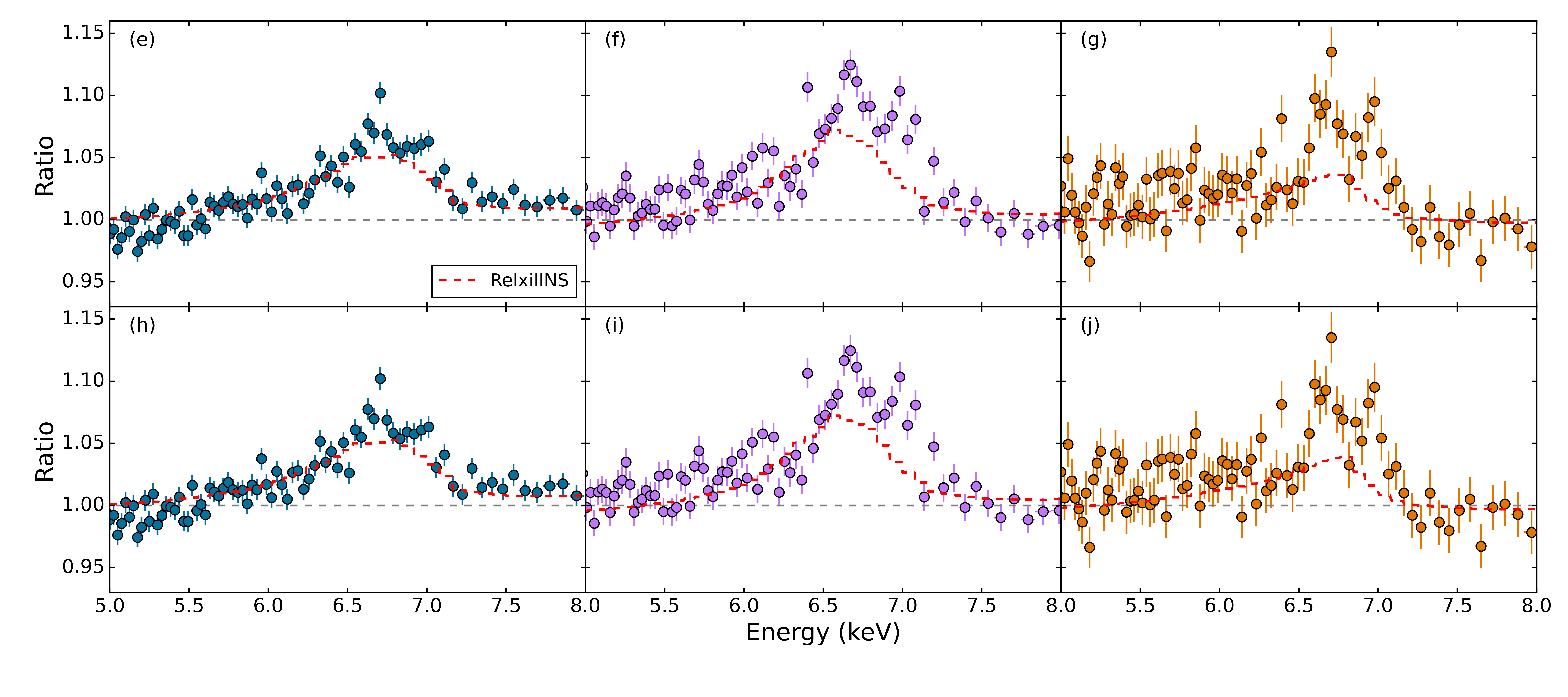}
 \end{center}
\caption{Panels (a) and (c) show the unfolded models and spectra in each branch for Model 1b and Model 2b as presented on Table~\ref{tab:reflection}, respectively. Model 1b assumes the Comptonization arises from the disk while Model 2b assumes the seed photons are a single-temperature blackbody. The data are color coded according to the spectral branch (NB: blue, SA: purple, and FB: orange). The disk component is denoted by the dot-dashed line, single-temperature blackbody by the dashed line, thermal Comptonization (i.e., {\sc thcomp*diskbb} or {\sc thcomp*bbody}) by the dot-dot-dot-dashed line, and reflection model by the solid line.  Panels (b) and (d) show the ratio of the data to the overall respective model. The inset in the lower panel shows the remaining residuals in the \xrism data after the reflection model is applied. There are narrow emission features at the $\sim5$\% level near 6.7 keV and 6.97 keV when using the reflection model alone regardless of the choice of seed photons for Comptonization. The residual at 6.7 keV improves with the addition of the plasma model component (see Figure 4). Panels (e)--(g) show the Fe line profile from applying Model 1a to the data as shown in Figure~\ref{fig:iron} with the reflection model predicted line component (red dashed line) in each branch as indicated in Table~\ref{tab:reflection}. For completeness, panels (h)--(j) show the Fe line profile when Model 2a is applied to the \xrism data and the reflection component from Model 2b is overlaid. There is no visual difference in the structure of the Fe line profile between continuum Model 1a and 2a. } {Alt text: Graphs showing the unfolded models and spectra, ratio of the model fit to the data, and iron lines with the model predicted profile of the \xrism/Resolve, \nustar, and \nicer data during different branches traced out by \source with subplots labeled a to j.}
\label{fig:relxillns}
\end{figure*}

Figure \ref{fig:iron} shows the Fe line profile of the \xrism/Resolve, \nicer, and \nustar data in the different branches when continuum Model 1a is applied. There is no visual difference in the Fe line region when Model 2a is applied instead (see Figure~\ref{fig:relxillns} panels e--j for comparison), therefore Figure \ref{fig:iron} is not dependent upon the choice of seed photons for Comptonization. Note that the short exposure times of the \nicer data in each branch as listed in Table~\ref{tab:obs} results in comparatively noisy data. A dual-peaked line is evident in all branches. We note that there is evidence of emission near 6.9 keV rather than an absorption line as reported in \cite{miller16} for the HB observation with {\it Chandra}.  Trying to model the apparent dual-peaked emission line in the NB to FB as a Gaussian absorption line near 6.8 keV with a width of $\sigma=0.05$ keV \citep{cackett10} superimposed on an emission line is unable to account for the observed line profile. Alternatively, the complex Fe line profile can be approximated with a combination of three Gaussian emission lines at fixed energy $E_{1}=6.4$ keV, $E_{2}=6.7$ keV, and $E_{3}=6.97$ keV with equivalent widths of 15--70 eV, 10--19 eV, and 2.0--3.7 eV, respectively, from the FB to NB.  The properties of the Gaussian lines are summarized in Table~\ref{tab:gauss} for the underlying continuum of Model 1a. The Gaussian line parameter values from using Model 2a all agree within uncertainty, but we only report the values from one underlying continuum description for clarity. The equivalent widths are similar to those reported for the Fe line in the HB using single Gaussian line components with lower energy resolution X-ray spectra \citep{dai09, cackett10, lamonaca24}.

\begin{table*}[t!]

\caption{Reflection and ionized plasma modeling of the joint \xrism, \nicer, and \nustar data in the normal branch (NB), soft apex (SA), and flaring branch (FB).}
\label{tab:apec} 

\begin{center}

\begin{tabular}{llccc}
\hline

Model & Parameter & \multicolumn{3}{c}{Model 1c} \\
& & NB & SA & FB \\
\hline

{\sc crabcor} 
& $C_{\rm FPMB}\ ^{\dagger}$ 
& $ 0.980 \pm0.002$
& $ 0.980 \pm0.002$
& $ 0.980 \pm0.002$

\\
& $C_{\rm XRISM}$
& $ 0.687 _{- 0.001 }^{+ 0.002 }$
& $ 0.766 \pm 0.002 $
& $ 0.670 _{- 0.002 }^{+ 0.003 }$

\\
& $C_{\rm NICER}$
& $ 0.744 \pm 0.004$
& $ 0.770 _{- 0.004 }^{+ 0.006 }$
& $ 0.701 _{- 0.004 }^{+ 0.009 }$

\\
& $\Delta \Gamma_{\rm XRISM}\ ^{\dagger}$ ($10^{-2}$)
& $ -0.24 \pm0.01$
& $ -0.24 \pm0.01$
& $ -0.24 \pm0.01$

\\

{\sc tbabs} 
&$N_{\mathrm{H}}\ ^{\dagger}$ ($10^{22}$ cm$^{-2}$) 
& $ 8.74 _{- 0.03 }^{+ 0.04 }$
& $ 8.74 _{- 0.03 }^{+ 0.04 }$
& $ 8.74 _{- 0.03 }^{+ 0.04 }$

\\

{\sc thcomp} 

& $kT_{e}$ (keV)
&  $ 2.91 \pm0.02$
&  $ 3.39 _{- 0.06 }^{+ 0.03 }$
&  $ 26 _{- 2 }^{+ 1 }$

\\
& cov\_frac 
&  $ 0.253 _{- 0.003 }^{+ 0.002 }$
&  $ 0.051 \pm0.003$
&  $ 0.0015 _{- 0.0002 }^{+ 0.0001 }$

\\

{\sc diskbb} 
& $kT_{\rm in}$ (keV) 
& $ 1.42 \pm 0.01$
& $ 1.17 \pm 0.01$
& $ 0.96 _{- 0.02 }^{+ 0.01 }$

\\

& norm$_{\rm disk}$ 
& $ 193 _{- 4 }^{+ 3 }$
& $ 328 _{- 7 }^{+ 12 }$
& $ 688 _{- 37 }^{+ 56 }$

\\

{\sc bbody} 
& $kT_{\rm bb}$ (keV) 
& $ 1.57 _{- 0.03 }^{+ 0.02 }$
& $ 1.49 \pm 0.01$
& $ 1.44 \pm 0.01$

\\
& norm$_{\rm bb}$ ($10^{-2}$)
& $ 1.9\pm 0.1$
& $ 6.1 \pm0.2$
& $ 12 \pm0.2$

\\

{\sc relxillNS} 
& $q$
& $ 2.24 _{- 0.07 }^{+ 0.05 }$
& $ 2.10 _{- 0.06 }^{+ 0.08 }$
& $ 2.26 _{- 0.04 }^{+ 0.08 }$

\\

& $i\ ^{\dagger}$ ($^{\circ}$) 
&$ 39 \pm 1$
&$ 39 \pm 1$
&$ 39 \pm 1$

\\

& $R_{\rm in}$ ($R_{\rm ISCO}$)
& $ 1.84 _{- 0.06 }^{+ 0.04 }$
& $ 1.5 \pm0.1$
& $ 1.6 \pm0.1$

\\

& $\log (\xi)$ 
& $ 2.1 \pm0.1$
& $ 1.98 _{- 0.04 }^{+ 0.05 }$
& $ 1.95 _{- 0.09 }^{+ 0.04 }$

\\

& $A_{\rm Fe}\ ^{\dagger}$ 
&  $ 9.9 _{- 0.3 }^{+ 0.1 }$
&  $ 9.9 _{- 0.3 }^{+ 0.1 }$
&  $ 9.9 _{- 0.3 }^{+ 0.1 }$

\\

& norm$_{\rm refl}$ ($10^{-3}$) 
& $ 1.81 _{- 0.08 }^{+ 0.06 }$
& $ 2.08 _{- 0.08 }^{+ 0.09 }$
& $ 1.3 \pm0.1$

\\

{\sc apec} 
& $kT$ (keV)
&$ 2.24 _{- 0.07 }^{+ 0.05 }$
&$ 1.95 _{- 0.09 }^{+ 0.04 }$
&$ 2.1 \pm0.1$

\\
& $A_{\rm Z}$ $^{\dagger}$
&$ 2.26 _{- 0.04 }^{+ 0.08 }$
&$ 2.26 _{- 0.04 }^{+ 0.08 }$
&$ 2.26 _{- 0.04 }^{+ 0.08 }$

\\
& norm$_{\rm apec}$ ($10^{-2}$)
&$ 2.4 \pm0.1$
&$ 2.4 \pm0.1$
&$ 6.9 _{- 0.3 }^{+ 0.4 }$

\\

\hline
\multicolumn{2}{c}{total fit statistic (dof)} & \multicolumn{3}{c}{ 14216.96 (10964)}\\ 
\hline

\multicolumn{4}{l}{$^{\dagger}=$ tied between all branches} 

\end{tabular}
\end{center}

\medskip
Note.---  Errors are reported at the 90\% confidence level.  Model 1c assumes the Comptonized seed photons arise from the disk. For \relxillns, the disk density is fixed at $\log(n_{e}/\rm cm^{-3})=19$, the dimensionless spin at $a=0$, and reflection fraction at $f_{refl}=-1$ so that the model returns the reflection spectrum only. For $a=0$, the value of $1\ R_{\rm ISCO}=6\ R_{g}$.

\end{table*}

\begin{figure*}[t!]
 \begin{center}
  \includegraphics[width=0.95\textwidth]{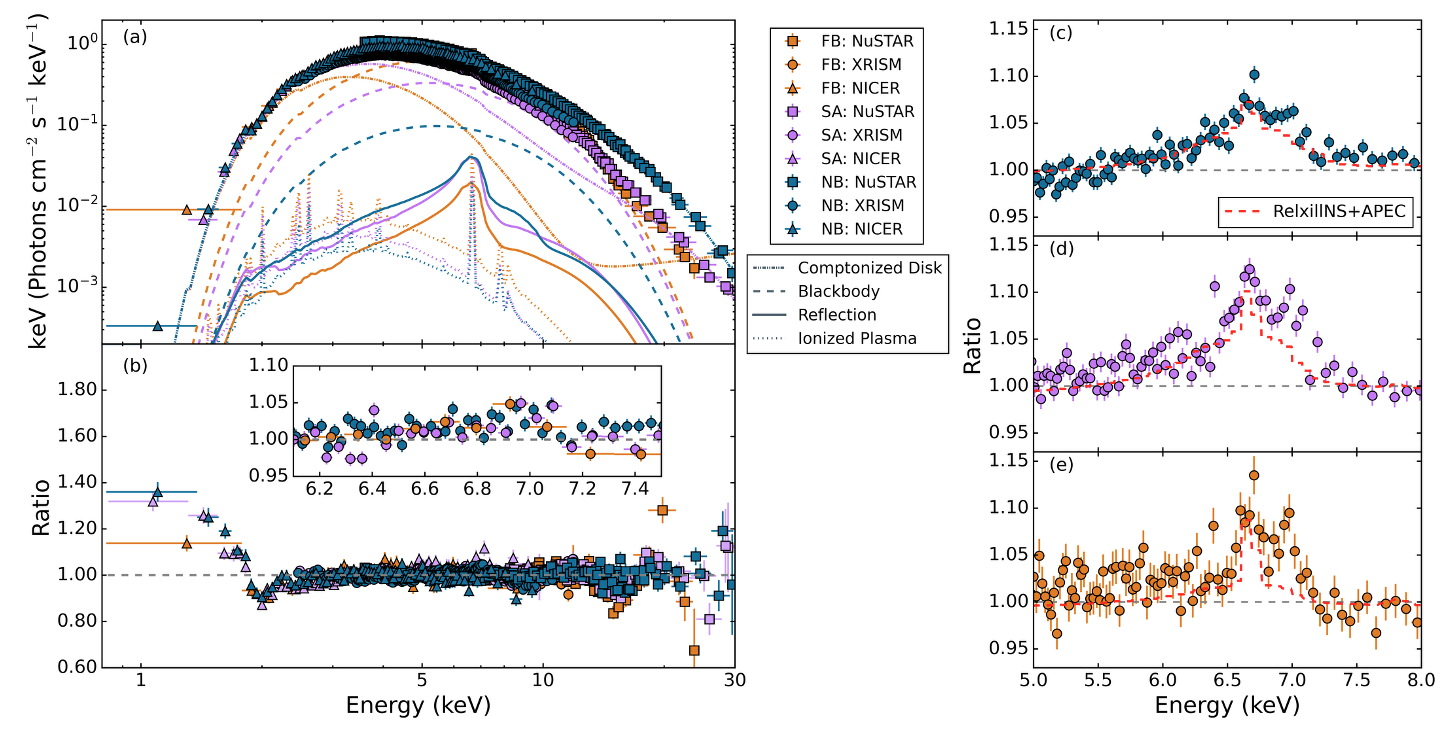} 
 \end{center}
\caption{Panel (a) shows the unfolded models and spectra in each branch for Model 1c as presented on Table~\ref{tab:apec}, respectively. Model 1c assumes the Comptonization arises from the disk. The data are color coded according to the spectral branch (NB: blue, SA: purple, and FB: orange). The disk component is denoted by the dot-dashed line, single-temperature blackbody by the dashed line, thermal Comptonization (i.e., {\sc thcomp*diskbb}) by the dot-dot-dot-dashed line, reflection model by the solid line, and ionized plasma component by the dotted line.  Panel (b) shows the ratio of the data to the overall respective model. The inset in the lower panel shows the remaining residuals in the \xrism data after the reflection and ionized plasma models are applied. The residual at 6.7 keV as shown in the inset ratios of Figure \ref{fig:relxillns} improves with the addition of the plasma model component. Panels (c)--(e) show the Fe line profile from applying Model 1a to the data as shown in Figure~\ref{fig:iron} with the combination of the reflection model and emission plasma (red dashed line) in each branch as indicated in Table~\ref{tab:apec}. } {Alt text: Graphs showing the unfolded models and spectra, ratio of the model fit to the data, and iron lines with the model predicted profile of the \xrism/Resolve, \nustar, and \nicer data during different branches traced out by \source with subplots labeled a to e.}
\label{fig:apec}
\end{figure*}

However, single emission line models are insufficient to describe the full reprocessed emission spectrum nor do they provide information regarding the physical properties of the emitting material.
Given our intent to explore if the structure in the line profile observed can be accounted for by reflection modeling, we apply \relxillns to model the reprocessed emission assuming the disk is illuminated by a boundary layer. As mentioned in \S~1, \relxillns has been able to model observed structure within the Fe line profile of the accreting NS LMXB Serpens X-1 for moderate energy resolution data \citep{ludlam18,ludlam24}. Other available reflection models that assume thermal illumination of the disk, such as {\sc reflionx\_bb} and {\sc bbrefl} \citep{ballantyne04}, which are both based on the parent code of {\sc reflionx} \citep{Ross2005}, do not have the same level of detailed treatment of atomic transitions that result in a lack of structure in the Fe line region when simulated at the resolution of \xrism/Resolve (see \citealt{Garcia22} for more details).  We do not test illumination by a Comptonized component as there is currently a lack of available self-consistent reflection models with a Comptonized input appropriate for NSs \citep{Ludlam19a}. Furthermore, the reflection convolution {\sc rfxconv} \citep{done06} that provides an approximation for Comptonized reflection is an interpolate between two reflection codes, {\sc reflionx} below 14 keV and {\sc pexriv} \citep{Magdziarz95} above 14 keV,  that has a fixed disk density of $10^{15}$ cm$^{-3}$, which is orders of magnitude below the expected density for NS LMXBs. Therefore, we focus solely on testing \relxillns and defer the use of other reflection models to a follow-up investigation.

For \relxillns, we set the reflection fraction to $-1$ so that only the reflection spectrum is returned by the model. The thermal illumination is tied to the blackbody temperature $kT_{bb}$ of the continuum. The inner and outer emissivity indices are tied together to create a single emissivity index, $q$, to describe the illumination profile. The break radius between the two emissivity indices is thus unused in the model. The outer disk radius is fixed at the maximum of 1000 \rg. The majority of NSs in LMXBs have a dimensionless spin parameter of $a\lesssim0.3$ \citep{galloway2008, miller2011}. Given that the spin frequency of \source has not been measured, $a$ is conservatively fixed at 0. This does not significantly impact our results as the difference in \risco between $a=0$ and $a=0.3$ is only $\sim1$~\rg. As mentioned in \S~\ref{sec:intro}, a standard thin accretion disk around a NS has a high disk density ($>10^{20}\ \rm cm^{-3}$: \citealt{shakura73, Frank02}), therefore we fix the disk density, $\log(n_{e}/\rm cm^{-3})$, at the maximum allowable value of 19. Though the impact of density on the reflection spectrum mainly changes the shape at lower energies \citep{Garcia22} that we are not as sensitive to with the \xrism gate valve being closed and the short exposure times of the \nicer spectra, this model limitation may have an impact on the inferred Fe abundance; see the discussion in \S \ref{sec:4}. The abundance of iron, $A_{\rm Fe}$, and inclination, $i (^{\circ})$, are free to vary but tied between the branches. Additionally, the inner disk radius, \rin, ionization parameter, $\log(\xi/\rm erg\ cm\ s^{-1})$, and normalization are left free to vary between branches. Results of applying the reflection model are presented in Table \ref{tab:reflection} and the unfolded model and spectra are shown in Figure \ref{fig:relxillns}.  Although the total fit statistic after the addition of reflection appears comparable between the two descriptions of thermal Comptonization, the $\Delta \rm BIC=11.22$ provides very strong evidence that Model 1b is preferred over Model 2b. Additionally, the lack of consistency between the inferred inner disk radius from the reflection model and emission radii from the continuum components in Model 2b provides further disfavors the model assuming the Comptonization arises from the single-temperature blackbody component (see \S\ref{sec:4}). Consequently, we do not consider the model further.

The reflection model does not fully describe the Fe line region. There are still narrow features 
that would require little to no broadening
near 6.7 keV and 6.97 keV at the $\sim5$\% (see inset panel in Figure \ref{fig:relxillns}b and \ref{fig:relxillns}d). Again, we note that the addition of a Gaussian absorption line near 6.8 keV does not remove these apparent narrow features. 
Given that \source belongs to the Cyg-like subclass of Z sources, these narrow emission features may be due to the presence of ionized plasma in the system similar to Cygnus X-2 \citep{Ludlam22}, the archetype for which the subclass name was given. 
Akin to the analysis of \mbox{Cygnus X-2}, we apply an ionized emission plasma model {\sc apec} \citep{Smith01} v.3.0.9 allowing for thermal broadening of lines. 
The temperature and normalization are free to vary, but the abundance of metals ($A_{\rm Z}$) in the plasma is tied between branches.  The results of the addition of the ionized plasma are presented in Table \ref{tab:apec}.
This provides a marginal reduction in the overall fit (total fit statistic decreased by 50.42 for 7 degrees of freedom), however $\Delta \rm BIC < 1$ in comparison to the Model 1b, which indicates weak to no evidence the addition of a single {\sc apec} component is statistically preferred over the reflection only fit.  While not statistically favored, this clearly improves the residuals in the Fe line profile near 6.7 keV even though a residual still remains near 6.97 keV (see inset in Figure~\ref{fig:apec}b). This may be indicative of the presence of a multi-temperature plasma or a product of the {\sc apec} model assuming a diffuse optically thin environment when material may be optically thick within the binary system (leading to optical depth effects that could potentially impact the absorption and scattering of the spectral features: \citealt{chakraborty21, chakraborty22}). However, exploration of these are outside the scope of this paper. A full analysis of the ionized plasma in the \xrism/Resolve spectra with photoionization code {\sc cloudy} is presented in P.\ Chakraborty et al.\ (2025, in prep.).

\section{Discussion}\label{sec:4}
We present the first spectral analysis of a 152 ks \xrism/Resolve observation of \source while the source was in the NB to FB. During the observation, simultaneous data were obtained with \nicer and \nustar for 2.7 ks and 22.47 ks, respectively. The source was observed at a 0.5--50 keV unabsorbed flux of $\sim2\times10^{-8}$ \fluxcgs. The spectra from each mission in each branch were modeled concurrently.  While this is not a formal cross-calibration study of \xrism with respect to the other two missions, which would require strict simultaneity of data considered, the \xrism/Resolve spectra are within $\pm1\%$ of the slope of \nustar. This is comparable to the cross-calibration of other established X-ray missions (see \citealt{madsen17}) and report of \xrism's performance at IACHEC\footnote{https://iachec.org/wp-content/presentations/2025/xrism1\_miller.pdf}.
The inferred column density of $N_{\rm H}\simeq8.7\times10^{22}\ \rm cm^{-2}$ of our analysis agrees well with the previously reported high column density along the line of sight \citep{lavagetto04, iaria06, dai09, cackett10}, but a deeper exploration of the ISM properties is presented in \cite{Corrales25}. 

The continuum was best described by a single-temperature blackbody from the NS or boundary layer and thermal Comptonization from the accretion disk (Model 1a,b,c). For posterity, we also considered Comptonization from the NS or boundary layer with emission from the accretion disk (Model 2a,b). While the overall fit statistic for the different seed photon inputs for the thermal Comptonization components were comparable, a statistical BIC test showed strong preference for Comptonization arising from the accretion disk, which we focus on for the remainder of the discussion. This is similar to \cite{bhargava23, bhargava24a, bhargava24b} which modeled the continuum of \source with a single-temperature blackbody, accretion disk, and Comptonization with seed photons input from the disk (though these analyses use the Comptonization code {\sc nthcomp} which is superseded by {\sc thcomp}) in the HB to NB. The electron temperature of the Comptonizing component in the NB reported in \cite{bhargava23,bhargava24b} ($kT_{e}\simeq2.9-3.1$) is consistent with our findings in the NB. The covering fraction of the thermal Comptonzing region decreases from the NB to the FB regardless of the input for the seed photons, which is consistent with the behavior reported in \cite{chattopadhyay24} and discussed further therein. The thermal components were hotter in the NB and cooled while expanding towards the FB in agreement with the expected behavior of Z sources throughout the Z track \citep{church10, Li25}. The unprecedented energy resolution of \xrism/Resolve reveals a complex structure in the Fe line region between 6--7 keV. We do not observe strong outflow signatures near 6.9 keV as was reported in \cite{miller16} for \chandra data while the source was in the HB, but we do not have \xrism/Resolve data while the source was in such a spectral state. The complex Fe line profile in the NB to FB could be described by the combination of three Gaussian emission lines with energies at 6.4~keV, 6.7~keV, and 6.97~keV, but does not fully encompass all the reprocessed emission within the spectra. 

In order to properly account for the reprocessed emission, we applied \relxillns, a self-consistent reflection model tailored for the illumination of the accretion disk by thermal emission from the NS or boundary layer. This is the first application of \relxillns to high energy resolution spectra of \source in the NB to FB.  The inclination inferred from the reflection model is $34^{\circ}$-- $41^{\circ}$, in agreement within the 90\% confidence level with previous moderate inclination reports \citep{fender00,dai09, cackett10, miller16, bhargava24a, bhargava24b, lamonaca24, Li25}.  The emissivity index, $q$, is shallower than the illumination expected for flat, Euclidean geometry ($q=3$), but agrees with the illumination profile for extended emission geometry around a slow spinning compact object \citep{kinch16,kinch19}
and the typical values of $1.5<q<4$ observed in NS LMXBs \citep{ludlam24}. The inner disk radius is truncated slightly outside the innermost stable circular orbit throughout each branch when considering the best fit model where Comptonization arises from the disk (Model 1b). This is aligned with the previous reports in \cite{dai09} and \cite{cackett10} of the relativistically broadened Fe line arising from a region of the disk close to the innermost stable circular orbit in the HB.
Furthermore, this is consistent with the values reported in \cite{lamonaca24} and \cite{Li25} for inclination and a slightly truncated accretion disk when using \relxillns for observations in the HB and throughout the Z track, respectively. We note that the model that assumes the Comptonizing component is fed by the NS or boundary layer (Model 2b) does return an inner disk radius consistent with the innermost stable circular orbit, however, this model is strongly disfavored over Model 1b as indicated through the statistical BIC test and these results are contrary to the existing literature. 

Furthermore, radius constraints can also be obtained from the normalizations of the continuum thermal components of Model 1b. Assuming a distance of $11\pm3$ kpc \citep{penninx93, fender00}, the standard color correction factor of 1.7 \citep{kubota01}, and inclination of $34^{\circ}\leq i\leq41^{\circ}$, then the inner disk radius from {\sc diskbb} when propagating errors ranges from 34.9--63.7 km, 45.8--84.4 km, and 68.2--127.4 km for the NB, SA, and FB, respectively. However, these values may be overestimated by up to a factor of 2.2 when considering zero-torque inner boundary conditions for thin disk accretion \citep{zimmerman05}. This would bring the values closer to the estimate from \relxillns, which corresponds to 17.4--24.8 km, 13.6--28.5 km, and 16.1--28.5 km from the NB to FB for a canonical NS mass of 1.4 \msun. We note that the mass of the NS in \source is currently unknown, so the conversion from \risco to km could also increase should the accreting NS be more massive than the canonically assumed value. For the spherical emission radius of the single-temperature blackbody component ({\sc bbody}), we find values of 10.8--19.3 km, 22.1--39.3 km, and 34.5--60.5 km from the NB to FB when assuming the same distance and color correction factor as used for the accretion disk component and propagating errors.
Each of these estimates rely on different assumptions but remain generally consistent with one another for Model 1b, whereas Model 2b exhibits a greater discrepancy between radius constraints; providing additional support of the choice of a Comptonized disk to obtain a coherent understanding of the emission geometry of \source.

The iron abundance is inferred to be supersolar (consistent with or near the upper limit of $10\times$ solar values). 
Supersolar abundances are known to occur when the reflection model attempts to mimic the spectral shape of a higher disk density spectra \citep{tomsick18, garcia18} and have been reported for other accreting Galactic X-ray binaries (e.g., 4U 1630-47: \citealt{Connors21}, Cygnus X-1: \citealt{tomsick18}, GS 1354-64: \citealt{liu23}, Serpens X-1: \citealt{ludlam18, hall25}, 4U 1636-53: \citealt{Ludlam17a}, 4U 1702-429: \citealt{Ludlam19a}). The disk density is anticipated to be in excess of $>10^{20}$ cm$^{-3}$ in the inner accretion disks around NSs. Given that the model has an upper limit of $10^{19}$ cm$^{-3}$, the supersolar iron abundances are not unexpected although they may well be unphysical. 
We test how the key results of the inner disk radius and inclination depend this near maximal inference of $A_{\rm Fe}$ by performing fits with the parameter fixed at lower value of $5\times$ solar abundance. In the case of $A_{\rm Fe}=5$, the total fit statistic increases by 22.6 for 1 degree of difference. The inclination is $40^{\circ} \pm 2^{\circ}$ and inner disk radii of $1.7 \pm 0.5$~\risco, $1.7 \pm 0.7$~\risco, and $1.6 \pm 0.2$~\risco for the NB to FB, respectively. While the overall fit statistic is worse when $A_{\rm Fe}$ is fixed in comparison to being left as a free parameter, the inclination and inner disk radii are consistent within the 90\% confidence level with values reported in Table \ref{tab:reflection} indicating that our results are not strongly biased by the near maximal supersolar abundance.
We note that the results of \relxillns on \source in the HB and throughout the Z-track as reported in \cite{lamonaca24} and \cite{Li25}, respectively, do not infer an iron abundance much higher than solar values, but this may be due to the lower energy-resolution spectra utilized therein that does not exhibit structure within the Fe line region to necessitate the model mimicking a higher disk density spectrum. Additionally, the ionization state of the disk is inferred to be moderate. By definition, the ionization parameter is directly proportional to the integrated illuminating flux ($F_{x}$) and inversely proportional to the disk density ($\xi=4\pi F_{x}/n_{e}$: \citealt{garcia13}), which therefore may not accurately represent the true ionization state of the disk in this case since we are limited to $n_{e}=10^{19}$ cm$^{-3}$. Modifications to extend the model to higher disk densities are underway and the impact on inferred parameters will be explored in a subsequent publication.

The reflection model encompasses the broad line component, but narrow emission line features remain near 6.7~keV and 6.97~keV. A recent spectral analysis of the Z source Cygnus X-2 using \nicer and \nustar data exhibited evidence of a 1~keV emission line from the presence of an ionized plasma in addition to reflection in the each branch \citep{Ludlam22}. Though the high column density along the line of sight prevents detection of a 1 keV line, \source is a Cyg-like Z source. Hence,
the narrow emission components in the Fe K region may arise from an ionized plasma similarly to Cygnus X-2. The addition of a plasma model to the spectra does reduce the appearance of the 6.7 keV narrow emission line (see Figure~\ref{fig:apec}). Similar to Cygnus X-2, the abundance of metals in the plasma is slightly higher than solar abundance, but not as inflated as the reflection model.  The temperature of the emitting region is cooler than or consistent with the electron temperature of the Comptonized continuum component. Conversely, the temperature of the ionized plasma in Cygnus X-2 was cooler than or comparable to the accretion disk \citep{Ludlam22}. This may indicate that the ionized plasma in \source is not a collisionally ionized plasma in the outer region of the accretion disk as in the case of Cygnus X-2 \citep{vrtilek86}, but rather could arise from photoionization of an extended accretion disk corona \citep{church06}. 
A quick calculation of the density of the plasma from the normalization of the {\sc apec} model returns a range of $2.1\times10^{15}$ cm$^{-3}$ -- $6.6\times10^{15}$ cm$^{-3}$ when propagating errors on the distance and normalization throughout the branches, as well as assuming that the electron and hydrogen densities are equivalent and uniformly distributed throughout the extended thin ($H/R<<1$) accretion disc corona with the typical radial extent $R\sim50000$~km for Z sources \citep{Church04} and height $H=0.01R$. This is the correct order of magnitude expected for an extended accretion disk corona that does not obscure the view to the central region in a Cyg-like Z source \citep{Jimenez02, schulz09}. 
A more in depth analysis of photoionization present in \source within the \xrism/Resolve spectra is presented in P.\ Chakraborty et al.\ (2025, in prep.).

The addition of the plasma model component does not significantly alter the results of the reflection model, but rather improves the constraints on the inner disk radius and inclination. This is likely due to the reflection model not needing to describe both the narrow and broad component in the Fe line region once the ionized plasma model is added. 
In this case, the inclination agrees within the 90\% confidence level ($38^{\circ}\leq i\leq40^{\circ}$) and the inner disk radius is slightly truncated prior to the innermost stable circular orbit in all branches. This is aligned with the previous reports in \cite{dai09} and \cite{cackett10} of the relativistically broadened Fe line arising from a region of the disk close to the innermost stable circular orbit in the HB.
Furthermore, this is consistent with the values reported in \cite{lamonaca24} and \cite{Li25} for inclination and a slightly truncated accretion disk when using \relxillns for observations throughout the Z track. 
A follow up analysis dividing the three branches into smaller increments in order to  map changes in the inner disk radius throughout the Z track is forthcoming. 

While the addition of an ionized plasma component improved modeling the structure within the emission line region similarly to Cygnus X-2, an updated reflection model with higher disk density with the most recent atomic data may provide further improvement. \cite{ding24} presented a recent extension of the standard flavor of {\sc relxill} (which assumes illumination by a cutoff power-law) to densities up to $10^{22}$ cm$^{-3}$ using the latest atomic data. Most notably, the updated reflection table with higher disk density plasma effects produce broader Fe line profiles for a given ionization state (see figure 2 in \citealt{ding24}) and stronger emission lines from highly ionized species like Fe {\sc xxv} at $\sim6.7$ keV. 
As mentioned previously, modifications to \relxillns are underway to extend the model to higher disk densities with the latest atomic data. The preliminary analysis of the high energy resolution \xrism/Resolve data of \source reveals a complex structure in the Fe line region, which is optimal for testing the upcoming expansion of the thermal reflection model. 

\section{Conclusion}
We present the first spectral analysis of a \xrism/Resolve observation of \source that was supplemented with contemporaneous \nicer and \nustar data. The source occupied the normal to flaring branch throughout the campaign. The superior energy resolution of \xrism/Resolve revealed that the broad Fe emission line contains a complex structure. We model the relativistic reprocessed emission using the publicly available, self-consistent reflection model \relxillns, which is specifically designed to model thermal illumination of accretion disks around NSs. This is the first time a full reflection model was used to describe the high energy resolution spectra from \source in the normal to flaring branch. A single relativistic reflection model was unable to adequately describe all the structure contained within the Fe line in the \xrism data. The addition of an ionized emitting plasma to the reflection modeling improved the fit in the Fe band and had properties consistent with those expected for an accretion disk corona, but still could not fully encompass all the emission of the Fe region. This suggests a more complex multi-temperature emitting plasma in the system or the need for a more sophisticated plasma code to properly describe the environmental effects impacting the emission. 
The addition of the ionized plasma to the reflection model fits did not significantly alter the results of the key reflection parameters such as the disk being truncated  
prior to the innermost stable circular orbit 
(in agreement with existing literature). However, updates to the reflection models currently in progress do predict large changes in parameter values for ionized disks such as seen here (especially the iron abundance, \citealt{ding24}). Future modeling with the updated reflection codes are essential for understanding the inner disk geometry, and in properly describing the complex Fe line structure from \source.

\begin{ack}
We thank the referee(s) for their comments that have helped to strengthen and clarify the results presented in this paper.
We thank Jeremy Hare (NASA/GSFC) for helpful discussion regarding the cross calibration of \nicer with \nustar. R.M.L.\ thanks Jon M.\ Miller (University of Michigan) for discussion of spectral modeling of \xrism/Resolve data. We acknowledge informative discussion of the \xrism calibration with Tahir Yaqoob (UMBC) and Misaki Mizumoto (University of Teacher Education Fukuoka). This research has made use of the NuSTAR Data Analysis Software (NuSTARDAS) jointly developed by the ASI Space Science Data Center (SSDC, Italy) and the California Institute of Technology (Caltech, USA). 

\end{ack}

\section*{Funding}
R.M.L.\ acknowledges support by NASA under award number 80NSSC23K0635. 
R.B.\ acknowledges support by NASA under award number 80GSFC21M0002. 
L.C.\ acknowledges support from NASA grants 80NSSC18K0978, 80NSSC20K0883, and 80NSSC25K7064. 
I.P.\ is supported by NASA through the Smithsonian Astrophysical Observatory (SAO) contract SV3-73016 to MIT for Support of the Chandra X-Ray Center (CXC) and Science Instruments. 
CXC is operated by SAO for and on behalf of NASA under contract NAS8-03060.
T.N.\ acknowledges the support by JSPS KAKENHI Grant
Numbers 23H05441 and 23K17695.

\section*{Data availability} 
The \nicer and \nustar data underlying this article are publicly available on the NASA data archive. The \xrism data are subject to a 12 months proprietary time from the completion of the \xrism Performance Verification phase data collection after which the data will be publicly available from NASA and JAXA online data archives.

\bibliographystyle{aasjournal.bst}
\bibliography{references}

\begin{thebibliography}{}
\expandafter\ifx\csname natexlab\endcsname\relax\def\natexlab#1{#1}\fi
\providecommand{\url}[1]{\href{#1}{#1}}
\providecommand{\dodoi}[1]{doi:~\href{http://doi.org/#1}{\nolinkurl{#1}}}
\providecommand{\doeprint}[1]{\href{http://ascl.net/#1}{\nolinkurl{http://ascl.net/#1}}}
\providecommand{\doarXiv}[1]{\href{https://arxiv.org/abs/#1}{\nolinkurl{https://arxiv.org/abs/#1}}}

\bibitem[{{Arnaud}(1996)}]{arnaud96}
{Arnaud}, K.~A. 1996, in Astronomical Society of the Pacific Conference Series, Vol. 101, Astronomical Data Analysis Software and Systems V, ed. G.~H. {Jacoby} \& J.~{Barnes}, 17

\bibitem[{{Ballantyne}(2004)}]{ballantyne04}
{Ballantyne}, D.~R. 2004, \mnras, 351, 57, \dodoi{10.1111/j.1365-2966.2004.07767.x}

\bibitem[{{Ba{\l}uci{\'n}ska-Church} {et~al.}(2010){Ba{\l}uci{\'n}ska-Church}, {Gibiec}, {Jackson}, \& {Church}}]{church10}
{Ba{\l}uci{\'n}ska-Church}, M., {Gibiec}, A., {Jackson}, N.~K., \& {Church}, M.~J. 2010, \aap, 512, A9, \dodoi{10.1051/0004-6361/200913199}

\bibitem[{{Bhargava} {et~al.}(2023){Bhargava}, {Bhattacharyya}, {Homan}, \& {Pahari}}]{bhargava23}
{Bhargava}, Y., {Bhattacharyya}, S., {Homan}, J., \& {Pahari}, M. 2023, \apj, 955, 102, \dodoi{10.3847/1538-4357/acee7a}

\bibitem[{{Bhargava} {et~al.}(2024{\natexlab{a}}){Bhargava}, {Ng}, {Zhang}, {Balasubramanian}, {Russell}, {Kaushik}, {Jadoliya}, {Ravi}, {Bhattacharyya}, {Pahari}, {Homan}, {Marshall}, {Chakrabarty}, \& {Carotenuto}}]{bhargava24a}
{Bhargava}, Y., {Ng}, M., {Zhang}, L., {et~al.} 2024{\natexlab{a}}, arXiv e-prints, arXiv:2405.19324, \dodoi{10.48550/arXiv.2405.19324}

\bibitem[{{Bhargava} {et~al.}(2024{\natexlab{b}}){Bhargava}, {Russell}, {Ng}, {Balasubramanian}, {Zhang}, {Ravi}, {Jadoliya}, {Bhattacharyya}, {Pahari}, {Homan}, {Marshall}, {Chakrabarty}, {Carotenuto}, \& {Kaushik}}]{bhargava24b}
{Bhargava}, Y., {Russell}, T.~D., {Ng}, M., {et~al.} 2024{\natexlab{b}}, arXiv e-prints, arXiv:2411.00350, \dodoi{10.48550/arXiv.2411.00350}

\bibitem[{{Bhattacharyya} \& {Strohmayer}(2007)}]{Bhattacharyya07}
{Bhattacharyya}, S., \& {Strohmayer}, T.~E. 2007, \apjl, 664, L103, \dodoi{10.1086/520844}

\bibitem[{{Bult} {et~al.}(2018){Bult}, {Altamirano}, {Arzoumanian}, {Cackett}, {Chakrabarty}, {Doty}, {Enoto}, {Gendreau}, {Guillot}, {Homan}, {Jaisawal}, {Lamb}, {Ludlam}, {Mahmoodifar}, {Markwardt}, {Okajima}, {Price}, {Strohmayer}, \& {Winternitz}}]{bult18}
{Bult}, P., {Altamirano}, D., {Arzoumanian}, Z., {et~al.} 2018, \apjl, 860, L9, \dodoi{10.3847/2041-8213/aac893}

\bibitem[{{Cackett} {et~al.}(2009){Cackett}, {Altamirano}, {Patruno}, {Miller}, {Reynolds}, {Linares}, \& {Wijnands}}]{cackett09}
{Cackett}, E.~M., {Altamirano}, D., {Patruno}, A., {et~al.} 2009, \apjl, 694, L21, \dodoi{10.1088/0004-637X/694/1/L21}

\bibitem[{{Cackett} {et~al.}(2008){Cackett}, {Miller}, {Bhattacharyya}, {Grindlay}, {Homan}, {van der Klis}, {Miller}, {Strohmayer}, \& {Wijnands}}]{cackett08}
{Cackett}, E.~M., {Miller}, J.~M., {Bhattacharyya}, S., {et~al.} 2008, \apj, 674, 415, \dodoi{10.1086/524936}

\bibitem[{{Cackett} {et~al.}(2010){Cackett}, {Miller}, {Ballantyne}, {Barret}, {Bhattacharyya}, {Boutelier}, {Miller}, {Strohmayer}, \& {Wijnands}}]{cackett10}
{Cackett}, E.~M., {Miller}, J.~M., {Ballantyne}, D.~R., {et~al.} 2010, \apj, 720, 205, \dodoi{10.1088/0004-637X/720/1/205}

\bibitem[{{Chakraborty} {et~al.}(2022){Chakraborty}, {Ferland}, {Chatzikos}, {Fabian}, {Bianchi}, {Guzm{\'a}n}, \& {Su}}]{chakraborty22}
{Chakraborty}, P., {Ferland}, G.~J., {Chatzikos}, M., {et~al.} 2022, \apj, 935, 70, \dodoi{10.3847/1538-4357/ac7eb9}

\bibitem[{{Chakraborty} {et~al.}(2021){Chakraborty}, {Ferland}, {Chatzikos}, {Guzm{\'a}n}, \& {Su}}]{chakraborty21}
{Chakraborty}, P., {Ferland}, G.~J., {Chatzikos}, M., {Guzm{\'a}n}, F., \& {Su}, Y. 2021, \apj, 912, 26, \dodoi{10.3847/1538-4357/abed4a}

\bibitem[{{Chattopadhyay} {et~al.}(2024){Chattopadhyay}, {Bhulla}, {Misra}, \& {Mandal}}]{chattopadhyay24}
{Chattopadhyay}, S., {Bhulla}, Y., {Misra}, R., \& {Mandal}, S. 2024, \mnras, 528, 6167, \dodoi{10.1093/mnras/stae389}

\bibitem[{{Church} \& {Ba{\l}uci{\'n}ska-Church}(2004)}]{Church04}
{Church}, M.~J., \& {Ba{\l}uci{\'n}ska-Church}, M. 2004, \mnras, 348, 955, \dodoi{10.1111/j.1365-2966.2004.07162.x}

\bibitem[{{Church} {et~al.}(2006){Church}, {Halai}, \& {Ba{\l}uci{\'n}ska-Church}}]{church06}
{Church}, M.~J., {Halai}, G.~S., \& {Ba{\l}uci{\'n}ska-Church}, M. 2006, \aap, 460, 233, \dodoi{10.1051/0004-6361:20065035}

\bibitem[{{Connors} {et~al.}(2021){Connors}, {Garc{\'\i}a}, {Tomsick}, {Hare}, {Dauser}, {Grinberg}, {Steiner}, {Mastroserio}, {Sridhar}, {Fabian}, {Jiang}, {Parker}, {Harrison}, \& {Kallman}}]{Connors21}
{Connors}, R. M.~T., {Garc{\'\i}a}, J.~A., {Tomsick}, J., {et~al.} 2021, \apj, 909, 146, \dodoi{10.3847/1538-4357/abdd2c}

\bibitem[{{Corrales} {et~al.}(2025){Corrales}, {Costantini}, {Zeeger}, {Gu}, {Takahashi}, {Moutard}, {Shidatsu}, {Miller}, {Mizumoto}, {Smith}, {Ballhausen}, {Chakraborty}, {Diaz Trigo}, {Ludlam}, {Nakagawa}, {Psaradaki}, {Yamada}, \& {Kilbourne}}]{Corrales25}
{Corrales}, L., {Costantini}, E., {Zeeger}, S., {et~al.} 2025, arXiv e-prints, arXiv:2506.08751, \dodoi{10.48550/arXiv.2506.08751}

\bibitem[{{D'A{\`\i}} {et~al.}(2009){D'A{\`\i}}, {Iaria}, {Di Salvo}, {Matt}, \& {Robba}}]{dai09}
{D'A{\`\i}}, A., {Iaria}, R., {Di Salvo}, T., {Matt}, G., \& {Robba}, N.~R. 2009, \apjl, 693, L1, \dodoi{10.1088/0004-637X/693/1/L1}

\bibitem[{{Dauser}(2010)}]{Dauser10}
{Dauser}, T. 2010, Master's thesis, Friedrich Alexander University of Erlangen-Nuremberg, Germany

\bibitem[{{Ding} {et~al.}(2024){Ding}, {Garc{\i}a}, {Kallman}, {Mendoza}, {Bautista}, {Harrison}, {Tomsick}, \& {Dong}}]{ding24}
{Ding}, Y., {Garc{\i}a}, J.~A., {Kallman}, T.~R., {et~al.} 2024, \apj, 974, 280, \dodoi{10.3847/1538-4357/ad76a1}

\bibitem[{{Done} \& {Gierli{\'n}ski}(2006)}]{done06}
{Done}, C., \& {Gierli{\'n}ski}, M. 2006, \mnras, 367, 659, \dodoi{10.1111/j.1365-2966.2005.09968.x}

\bibitem[{{Fabian} {et~al.}(2000){Fabian}, {Iwasawa}, {Reynolds}, \& {Young}}]{Fabian00}
{Fabian}, A.~C., {Iwasawa}, K., {Reynolds}, C.~S., \& {Young}, A.~J. 2000, \pasp, 112, 1145, \dodoi{10.1086/316610}

\bibitem[{{Fender} \& {Hendry}(2000)}]{fender00}
{Fender}, R.~P., \& {Hendry}, M.~A. 2000, \mnras, 317, 1, \dodoi{10.1046/j.1365-8711.2000.03443.x}

\bibitem[{{Frank} {et~al.}(2002){Frank}, {King}, \& {Raine}}]{Frank02}
{Frank}, J., {King}, A., \& {Raine}, D.~J. 2002, {Accretion Power in Astrophysics: Third Edition}

\bibitem[{{Galloway} {et~al.}(2008){Galloway}, {Muno}, {Hartman}, {Psaltis}, \& {Chakrabarty}}]{galloway2008}
{Galloway}, D.~K., {Muno}, M.~P., {Hartman}, J.~M., {Psaltis}, D., \& {Chakrabarty}, D. 2008, \apjs, 179, 360, \dodoi{10.1086/592044}

\bibitem[{{Garc{\'\i}a} {et~al.}(2013){Garc{\'\i}a}, {Dauser}, {Reynolds}, {Kallman}, {McClintock}, {Wilms}, \& {Eikmann}}]{garcia13}
{Garc{\'\i}a}, J., {Dauser}, T., {Reynolds}, C.~S., {et~al.} 2013, \apj, 768, 146, \dodoi{10.1088/0004-637X/768/2/146}

\bibitem[{{Garc{\'\i}a} {et~al.}(2014){Garc{\'\i}a}, {Dauser}, {Lohfink}, {Kallman}, {Steiner}, {McClintock}, {Brenneman}, {Wilms}, {Eikmann}, {Reynolds}, \& {Tombesi}}]{garcia14}
{Garc{\'\i}a}, J., {Dauser}, T., {Lohfink}, A., {et~al.} 2014, \apj, 782, 76, \dodoi{10.1088/0004-637X/782/2/76}

\bibitem[{{Garc{\'\i}a} {et~al.}(2022){Garc{\'\i}a}, {Dauser}, {Ludlam}, {Parker}, {Fabian}, {Harrison}, \& {Wilms}}]{Garcia22}
{Garc{\'\i}a}, J.~A., {Dauser}, T., {Ludlam}, R., {et~al.} 2022, \apj, 926, 13, \dodoi{10.3847/1538-4357/ac3cb7}

\bibitem[{{Garc{\'\i}a} {et~al.}(2018){Garc{\'\i}a}, {Kallman}, {Bautista}, {Mendoza}, {Deprince}, {Palmeri}, \& {Quinet}}]{garcia18}
{Garc{\'\i}a}, J.~A., {Kallman}, T.~R., {Bautista}, M., {et~al.} 2018, in Astronomical Society of the Pacific Conference Series, Vol. 515, Workshop on Astrophysical Opacities, 282, \dodoi{10.48550/arXiv.1805.00581}

\bibitem[{{Gendreau} {et~al.}(2012){Gendreau}, {Arzoumanian}, \& {Okajima}}]{Gendreau12}
{Gendreau}, K.~C., {Arzoumanian}, Z., \& {Okajima}, T. 2012, in Society of Photo-Optical Instrumentation Engineers (SPIE) Conference Series, Vol. 8443, Space Telescopes and Instrumentation 2012: Ultraviolet to Gamma Ray, ed. T.~{Takahashi}, S.~S. {Murray}, \& J.-W.~A. {den Herder} (Bellingham, WA: SPIE), 844313, \dodoi{10.1117/12.926396}

\bibitem[{{Hall} {et~al.}(2025){Hall}, {Ludlam}, {Miller}, {Fabian}, {Tomsick}, {Coley}, {Garc{\'\i}a}, \& {Coughenour}}]{hall25}
{Hall}, H., {Ludlam}, R.~M., {Miller}, J.~M., {et~al.} 2025, \apj, 980, 234, \dodoi{10.3847/1538-4357/adaeaa}

\bibitem[{{Harrison} {et~al.}(2013){Harrison}, {Craig}, {Christensen}, {Hailey}, {Zhang}, {Boggs}, {Stern}, {Cook}, {Forster}, {Giommi}, {Grefenstette}, {Kim}, {Kitaguchi}, {Koglin}, {Madsen}, {Mao}, {Miyasaka}, {Mori}, {Perri}, {Pivovaroff}, {Puccetti}, {Rana}, {Westergaard}, {Willis}, {Zoglauer}, {An}, {Bachetti}, {Barri{\`e}re}, {Bellm}, {Bhalerao}, {Brejnholt}, {Fuerst}, {Liebe}, {Markwardt}, {Nynka}, {Vogel}, {Walton}, {Wik}, {Alexander}, {Cominsky}, {Hornschemeier}, {Hornstrup}, {Kaspi}, {Madejski}, {Matt}, {Molendi}, {Smith}, {Tomsick}, {Ajello}, {Ballantyne}, {Balokovi{\'c}}, {Barret}, {Bauer}, {Blandford}, {Brandt}, {Brenneman}, {Chiang}, {Chakrabarty}, {Chenevez}, {Comastri}, {Dufour}, {Elvis}, {Fabian}, {Farrah}, {Fryer}, {Gotthelf}, {Grindlay}, {Helfand}, {Krivonos}, {Meier}, {Miller}, {Natalucci}, {Ogle}, {Ofek}, {Ptak}, {Reynolds}, {Rigby}, {Tagliaferri}, {Thorsett}, {Treister}, \& {Urry}}]{harrison13}
{Harrison}, F.~A., {Craig}, W.~W., {Christensen}, F.~E., {et~al.} 2013, \apj, 770, 103, \dodoi{10.1088/0004-637X/770/2/103}

\bibitem[{{Hasinger} \& {van der Klis}(1989)}]{hasinger89}
{Hasinger}, G., \& {van der Klis}, M. 1989, \aap, 225, 79

\bibitem[{{Hasinger} {et~al.}(1990){Hasinger}, {van der Klis}, {Ebisawa}, {Dotani}, \& {Mitsuda}}]{hasinger90}
{Hasinger}, G., {van der Klis}, M., {Ebisawa}, K., {Dotani}, T., \& {Mitsuda}, K. 1990, \aap, 235, 131

\bibitem[{{Homan} {et~al.}(2007){Homan}, {van der Klis}, {Wijnands}, {Belloni}, {Fender}, {Klein-Wolt}, {Casella}, {M{\'e}ndez}, {Gallo}, {Lewin}, \& {Gehrels}}]{homan07}
{Homan}, J., {van der Klis}, M., {Wijnands}, R., {et~al.} 2007, \apj, 656, 420, \dodoi{10.1086/510447}

\bibitem[{{Homan} {et~al.}(2010){Homan}, {van der Klis}, {Fridriksson}, {Remillard}, {Wijnands}, {M{\'e}ndez}, {Lin}, {Altamirano}, {Casella}, {Belloni}, \& {Lewin}}]{homan10}
{Homan}, J., {van der Klis}, M., {Fridriksson}, J.~K., {et~al.} 2010, \apj, 719, 201, \dodoi{10.1088/0004-637X/719/1/201}

\bibitem[{{Iaria} {et~al.}(2006){Iaria}, {Lavagetto}, {Di Salvo}, {D'A{\`\i}}, {Burderi}, {Stella}, \& {Robba}}]{iaria06}
{Iaria}, R., {Lavagetto}, G., {Di Salvo}, T., {et~al.} 2006, Chinese Journal of Astronomy and Astrophysics Supplement, 6, 257, \dodoi{10.1088/1009-9271/6/S1/33}

\bibitem[{{Ibragimov} \& {Poutanen}(2009)}]{ibragimov2009}
{Ibragimov}, A., \& {Poutanen}, J. 2009, \mnras, 400, 492, \dodoi{10.1111/j.1365-2966.2009.15477.x}

\bibitem[{{Inogamov} \& {Sunyaev}(1999)}]{inogamov99}
{Inogamov}, N.~A., \& {Sunyaev}, R.~A. 1999, Astronomy Letters, 25, 269, \dodoi{10.48550/arXiv.astro-ph/9904333}

\bibitem[{{Ishisaki} {et~al.}(2018){Ishisaki}, {Yamada}, {Seta}, {Tashiro}, {Takeda}, {Terada}, {Kato}, {Tsujimoto}, {Koyama}, {Mitsuda}, {Sawada}, {Boyce}, {Chiao}, {Watanabe}, {Leutenegger}, {Eckart}, {Porter}, \& {Kilbourne}}]{ishisaki18}
{Ishisaki}, Y., {Yamada}, S., {Seta}, H., {et~al.} 2018, Journal of Astronomical Telescopes, Instruments, and Systems, 4, 011217, \dodoi{10.1117/1.JATIS.4.1.011217}

\bibitem[{{Jimenez-Garate} {et~al.}(2002){Jimenez-Garate}, {Raymond}, \& {Liedahl}}]{Jimenez02}
{Jimenez-Garate}, M.~A., {Raymond}, J.~C., \& {Liedahl}, D.~A. 2002, \apj, 581, 1297, \dodoi{10.1086/344364}

\bibitem[{{Jonker} {et~al.}(1998){Jonker}, {Wijnands}, {van der Klis}, {Psaltis}, {Kuulkers}, \& {Lamb}}]{jonker98}
{Jonker}, P.~G., {Wijnands}, R., {van der Klis}, M., {et~al.} 1998, \apjl, 499, L191, \dodoi{10.1086/311372}

\bibitem[{{Kaastra} \& {Bleeker}(2016)}]{kb16}
{Kaastra}, J.~S., \& {Bleeker}, J.~A.~M. 2016, \aap, 587, A151, \dodoi{10.1051/0004-6361/201527395}

\bibitem[{{Kinch} {et~al.}(2016){Kinch}, {Schnittman}, {Kallman}, \& {Krolik}}]{kinch16}
{Kinch}, B.~E., {Schnittman}, J.~D., {Kallman}, T.~R., \& {Krolik}, J.~H. 2016, \apj, 826, 52, \dodoi{10.3847/0004-637X/826/1/52}

\bibitem[{{Kinch} {et~al.}(2019){Kinch}, {Schnittman}, {Kallman}, \& {Krolik}}]{kinch19}
---. 2019, \apj, 873, 71, \dodoi{10.3847/1538-4357/ab05d5}

\bibitem[{{King} {et~al.}(2016){King}, {Tomsick}, {Miller}, {Chenevez}, {Barret}, {Boggs}, {Chakrabarty}, {Christensen}, {Craig}, {F{\"u}rst}, {Hailey}, {Harrison}, {Parker}, {Stern}, {Romano}, {Walton}, \& {Zhang}}]{king16}
{King}, A.~L., {Tomsick}, J.~A., {Miller}, J.~M., {et~al.} 2016, \apjl, 819, L29, \dodoi{10.3847/2041-8205/819/2/L29}

\bibitem[{{Kubota} {et~al.}(2001){Kubota}, {Makishima}, \& {Ebisawa}}]{kubota01}
{Kubota}, A., {Makishima}, K., \& {Ebisawa}, K. 2001, \apjl, 560, L147, \dodoi{10.1086/324377}

\bibitem[{{Kuulkers} {et~al.}(1997){Kuulkers}, {van der Klis}, {Oosterbroek}, {van Paradijs}, \& {Lewin}}]{kuulkers97}
{Kuulkers}, E., {van der Klis}, M., {Oosterbroek}, T., {van Paradijs}, J., \& {Lewin}, W.~H.~G. 1997, \mnras, 287, 495, \dodoi{10.1093/mnras/287.3.495}

\bibitem[{{La Monaca} {et~al.}(2024){La Monaca}, {Di Marco}, {Ludlam}, {Bobrikova}, {Poutanen}, {Li}, \& {Xie}}]{lamonaca24}
{La Monaca}, F., {Di Marco}, A., {Ludlam}, R.~M., {et~al.} 2024, arXiv e-prints, arXiv:2410.00972, \dodoi{10.48550/arXiv.2410.00972}

\bibitem[{{Lavagetto} {et~al.}(2004){Lavagetto}, {Iaria}, {di Salvo}, {Burderi}, {Robba}, {Frontera}, \& {Stella}}]{lavagetto04}
{Lavagetto}, G., {Iaria}, R., {di Salvo}, T., {et~al.} 2004, Nuclear Physics B Proceedings Supplements, 132, 616, \dodoi{10.1016/j.nuclphysbps.2004.04.106}

\bibitem[{{Li} {et~al.}(2025){Li}, {Ludlam}, {Buisson}, {Sudha}, {Rossland}, {Mastroserio}, {Brumback}, {Garc\textbackslash'ia}, {Grefenstette}, {La Monaca}, {Saavedra}, \& {Di Marco}}]{Li25}
{Li}, S., {Ludlam}, R.~M., {Buisson}, D.~J.~K., {et~al.} 2025, arXiv e-prints, arXiv:2503.20050, \dodoi{10.48550/arXiv.2503.20050}

\bibitem[{{Lin} {et~al.}(2009){Lin}, {Remillard}, \& {Homan}}]{lin09}
{Lin}, D., {Remillard}, R.~A., \& {Homan}, J. 2009, \apj, 696, 1257, \dodoi{10.1088/0004-637X/696/2/1257}

\bibitem[{{Liu} {et~al.}(2023){Liu}, {Jiang}, {Zhang}, {Bambi}, {Fabian}, {Garc{\'\i}a}, {Ingram}, {Kara}, {Steiner}, {Tomsick}, {Walton}, \& {Young}}]{liu23}
{Liu}, H., {Jiang}, J., {Zhang}, Z., {et~al.} 2023, \apj, 951, 145, \dodoi{10.3847/1538-4357/acd8b9}

\bibitem[{{Ludlam}(2024)}]{ludlam24}
{Ludlam}, R.~M. 2024, \apss, 369, 16, \dodoi{10.1007/s10509-024-04281-y}

\bibitem[{{Ludlam} {et~al.}(2017{\natexlab{a}}){Ludlam}, {Miller}, {Degenaar}, {Sanna}, {Cackett}, {Altamirano}, \& {King}}]{Ludlam17c}
{Ludlam}, R.~M., {Miller}, J.~M., {Degenaar}, N., {et~al.} 2017{\natexlab{a}}, \apj, 847, 135, \dodoi{10.3847/1538-4357/aa8b1b}

\bibitem[{{Ludlam} {et~al.}(2017{\natexlab{b}}){Ludlam}, {Miller}, {Bachetti}, {Barret}, {Bostrom}, {Cackett}, {Degenaar}, {Di Salvo}, {Natalucci}, {Tomsick}, {Paerels}, \& {Parker}}]{Ludlam17a}
{Ludlam}, R.~M., {Miller}, J.~M., {Bachetti}, M., {et~al.} 2017{\natexlab{b}}, \apj, 836, 140, \dodoi{10.3847/1538-4357/836/1/140}

\bibitem[{{Ludlam} {et~al.}(2018){Ludlam}, {Miller}, {Arzoumanian}, {Bult}, {Cackett}, {Chakrabarty}, {Dauser}, {Enoto}, {Fabian}, {Garc{\'\i}a}, {Gendreau}, {Guillot}, {Homan}, {Jaisawal}, {Keek}, {La Marr}, {Malacaria}, {Markwardt}, {Steiner}, \& {Strohmayer}}]{ludlam18}
{Ludlam}, R.~M., {Miller}, J.~M., {Arzoumanian}, Z., {et~al.} 2018, \apjl, 858, L5, \dodoi{10.3847/2041-8213/aabee6}

\bibitem[{{Ludlam} {et~al.}(2019){Ludlam}, {Miller}, {Barret}, {Cackett}, {Coughenour}, {Dauser}, {Degenaar}, {Garc{\'\i}a}, {Harrison}, \& {Paerels}}]{Ludlam19a}
{Ludlam}, R.~M., {Miller}, J.~M., {Barret}, D., {et~al.} 2019, \apj, 873, 99, \dodoi{10.3847/1538-4357/ab0414}

\bibitem[{{Ludlam} {et~al.}(2022){Ludlam}, {Cackett}, {Garc{\'\i}a}, {Miller}, {Stevens}, {Fabian}, {Homan}, {Ng}, {Guillot}, {Buisson}, \& {Chakrabarty}}]{Ludlam22}
{Ludlam}, R.~M., {Cackett}, E.~M., {Garc{\'\i}a}, J.~A., {et~al.} 2022, \apj, 927, 112, \dodoi{10.3847/1538-4357/ac5028}

\bibitem[{{Madsen} {et~al.}(2017){Madsen}, {Beardmore}, {Forster}, {Guainazzi}, {Marshall}, {Miller}, {Page}, \& {Stuhlinger}}]{madsen17}
{Madsen}, K.~K., {Beardmore}, A.~P., {Forster}, K., {et~al.} 2017, \aj, 153, 2, \dodoi{10.3847/1538-3881/153/1/2}

\bibitem[{{Magdziarz} \& {Zdziarski}(1995)}]{Magdziarz95}
{Magdziarz}, P., \& {Zdziarski}, A.~A. 1995, \mnras, 273, 837, \dodoi{10.1093/mnras/273.3.837}

\bibitem[{{Miller} {et~al.}(2011){Miller}, {Maitra}, {Cackett}, {Bhattacharyya}, \& {Strohmayer}}]{miller2011}
{Miller}, J.~M., {Maitra}, D., {Cackett}, E.~M., {Bhattacharyya}, S., \& {Strohmayer}, T.~E. 2011, \apjl, 731, L7, \dodoi{10.1088/2041-8205/731/1/L7}

\bibitem[{{Miller} {et~al.}(2016){Miller}, {Raymond}, {Cackett}, {Grinberg}, \& {Nowak}}]{miller16}
{Miller}, J.~M., {Raymond}, J., {Cackett}, E., {Grinberg}, V., \& {Nowak}, M. 2016, \apjl, 822, L18, \dodoi{10.3847/2041-8205/822/1/L18}

\bibitem[{{Mitsuda} {et~al.}(1989){Mitsuda}, {Inoue}, {Nakamura}, \& {Tanaka}}]{mitsuda89}
{Mitsuda}, K., {Inoue}, H., {Nakamura}, N., \& {Tanaka}, Y. 1989, \pasj, 41, 97

\bibitem[{{Mizumoto} {et~al.}(2025a){Mizumoto}, {Tamba}, {Tsujimoto}, {Cumbee}, {Eckart}, {Hodges-Kluck}, {Ishisaki}, {Kilbourne}, {Leutenegger}, {Porter}, {Sawada}, {Takei}, {Uchida}, \& {Yamada}}]{mizumoto25}
{Mizumoto}, M., {Tamba}, T., {Tsujimoto}, M., {et~al.} 2025a, arXiv e-prints, arXiv:2501.03283, \dodoi{10.48550/arXiv.2501.03283}

\bibitem[{{Mizumoto} {et~al.}(2025b){Mizumoto}, {Kanemaru}, {Yamada}, {Kilbourne}, {Eckart}, {Hodges-Kluck}, {Ishisaki}, {Porter}, {Pottschmidt}, \& {Tamba}}]{2025arXiv250606692M}
{Mizumoto}, M., {Kanemaru}, Y., {Yamada}, S., {et~al.} 2025b, arXiv e-prints, arXiv:2506.06692.
\newblock \doarXiv{2506.06692}

\bibitem[{{Moutard} {et~al.}(2023){Moutard}, {Ludlam}, {Garc{\'\i}a}, {Altamirano}, {Buisson}, {Cackett}, {Chenevez}, {Degenaar}, {Fabian}, {Homan}, {Jaodand}, {Pike}, {Shaw}, {Strohmayer}, {Tomsick}, \& {Coughenour}}]{moutard23}
{Moutard}, D.~L., {Ludlam}, R.~M., {Garc{\'\i}a}, J.~A., {et~al.} 2023, \apj, 957, 27, \dodoi{10.3847/1538-4357/acf4f3}

\bibitem[{{Ng} {et~al.}(2023){Ng}, {Hughes}, {Homan}, {Miller}, {Pike}, {Altamirano}, {Bult}, {Chakrabarty}, {Buisson}, {Coughenour}, {Fender}, {Guillot}, {G{\"u}ver}, {Jaisawal}, {Jaodand}, {Malacaria}, {Miller-Jones}, {Sanna}, {Sivakoff}, {Strohmayer}, {Tomsick}, \& {van den Eijnden}}]{Ng2023}
{Ng}, M., {Hughes}, A.~K., {Homan}, J., {et~al.} 2023, arXiv e-prints, arXiv:2310.01511, \dodoi{10.48550/arXiv.2310.01511}

\bibitem[{{Papitto} {et~al.}(2009){Papitto}, {Di Salvo}, {D'A{\`\i}}, {Iaria}, {Burderi}, {Riggio}, {Menna}, \& {Robba}}]{papitto09}
{Papitto}, A., {Di Salvo}, T., {D'A{\`\i}}, A., {et~al.} 2009, \aap, 493, L39, \dodoi{10.1051/0004-6361:200811401}

\bibitem[{{Penninx} {et~al.}(1993){Penninx}, {Zwarthoed}, {van Paradijs}, {van der Klis}, {Lewin}, \& {Dotani}}]{penninx93}
{Penninx}, W., {Zwarthoed}, G.~A.~A., {van Paradijs}, J., {et~al.} 1993, \aap, 267, 92

\bibitem[{{Popham} \& {Sunyaev}(2001)}]{popham2001}
{Popham}, R., \& {Sunyaev}, R. 2001, \apj, 547, 355, \dodoi{10.1086/318336}

\bibitem[{{Ross} \& {Fabian}(2005)}]{Ross2005}
{Ross}, R.~R., \& {Fabian}, A.~C. 2005, \mnras, 358, 211, \dodoi{10.1111/j.1365-2966.2005.08797.x}

\bibitem[{{Schulz} {et~al.}(2009){Schulz}, {Huenemoerder}, {Ji}, {Nowak}, {Yao}, \& {Canizares}}]{schulz09}
{Schulz}, N.~S., {Huenemoerder}, D.~P., {Ji}, L., {et~al.} 2009, \apjl, 692, L80, \dodoi{10.1088/0004-637X/692/2/L80}

\bibitem[{{Shakura} \& {Sunyaev}(1973)}]{shakura73}
{Shakura}, N.~I., \& {Sunyaev}, R.~A. 1973, \aap, 24, 337

\bibitem[{{Smith} {et~al.}(2001){Smith}, {Brickhouse}, {Liedahl}, \& {Raymond}}]{Smith01}
{Smith}, R.~K., {Brickhouse}, N.~S., {Liedahl}, D.~A., \& {Raymond}, J.~C. 2001, \apjl, 556, L91, \dodoi{10.1086/322992}

\bibitem[{{Steiner} {et~al.}(2010){Steiner}, {McClintock}, {Remillard}, {Gou}, {Yamada}, \& {Narayan}}]{steiner10}
{Steiner}, J.~F., {McClintock}, J.~E., {Remillard}, R.~A., {et~al.} 2010, \apjl, 718, L117, \dodoi{10.1088/2041-8205/718/2/L117}

\bibitem[{{Sunyaev} {et~al.}(1991){Sunyaev}, {Arefev}, {Borozdin}, {Gilfanov}, {Efremov}, {Kaniovskii}, {Churazov}, {Kendziorra}, {Mony}, {Kretschmar}, {Maisack}, {Staubert}, {Dobereiner}, {Englhauser}, {Pietsch}, {Reppin}, {Trumper}, {Skinner}, {Nottingham}, {Pan}, \& {Willmore}}]{Sunyaev1991}
{Sunyaev}, R.~A., {Arefev}, V.~A., {Borozdin}, K.~N., {et~al.} 1991, Soviet Astronomy Letters, 17, 409

\bibitem[{{Tashiro} {et~al.}(2018){Tashiro}, {Maejima}, {Toda}, {Kelley}, {Reichenthal}, {Lobell}, {Petre}, {Guainazzi}, {Costantini}, {Edison}, {Fujimoto}, {Grim}, {Hayashida}, {den Herder}, {Ishisaki}, {Paltani}, {Matsushita}, {Mori}, {Sneiderman}, {Takei}, {Terada}, {Tomida}, {Akamatsu}, {Angelini}, {Arai}, {Awaki}, {Babyk}, {Bamba}, {Barfknecht}, {Barnstable}, {Bialas}, {Blagojevic}, {Bonafede}, {Brambora}, {Brenneman}, {Brown}, {Brown}, {Burns}, {Canavan}, {Carnahan}, {Chiao}, {Comber}, {Corrales}, {de Vries}, {Dercksen}, {Diaz-Trigo}, {Dillard}, {DiPirro}, {Done}, {Dotani}, {Ebisawa}, {Eckart}, {Enoto}, {Ezoe}, {Ferrigno}, {Fukazawa}, {Fujita}, {Furuzawa}, {Gallo}, {Graham}, {Gu}, {Hagino}, {Hamaguchi}, {Hatsukade}, {Hawes}, {Hayashi}, {Hegarty}, {Hell}, {Hiraga}, {Hodges-Kluck}, {Holland}, {Hornschemeier}, {Hoshino}, {Ichinohe}, {Iizuka}, {Ishibashi}, {Ishida}, {Ishikawa}, {Ishimura}, {James}, {Kallman}, {Kara}, {Katsuda}, {Kenyon}, {Kilbourne}, {Kimball}, {Kitaguti}, {Kitamoto}, {Kobayashi},
  {Kohmura}, {Koyama}, {Kubota}, {Leutenegger}, {Lockard}, {Loewenstein}, {Maeda}, {Marbley}, {Markevitch}, {Matsumoto}, {Matsuzaki}, {McCammon}, {McNamara}, {Miko}, {Miller}, {Miller}, {Minesugi}, {Mitsuishi}, {Mizuno}, {Mori}, {Mukai}, {Murakami}, {Mushotzky}, {Nakajima}, {Nakamura}, {Nakashima}, {Nakazawa}, {Natsukari}, {Nigo}, {Nishioka}, {Nobukawa}, {Nobukawa}, {Noda}, {Odaka}, {Ogawa}, {Ohashi}, {Ohno}, {Ohta}, {Okajima}, {Okamoto}, {Onizuka}, {Ota}, {Ozaki}, {Plucinsky}, {Porter}, {Pottschmidt}, {Sato}, {Sato}, {Sawada}, {Seta}, {Shelton}, {Shibano}, {Shida}, {Shidatsu}, {Shirron}, {Simionescu}, {Smith}, {Someya}, {Soong}, {Suagawara}, {Szymkowiak}, {Takahashi}, {Tamagawa}, {Tamura}, {Tanaka}, {Terashima}, {Tsuboi}, {Tsujimoto}, {Tsunemi}, {Tsuru}, {Uchida}, {Uchiyama}, {Ueda}, {Uno}, {Walsh}, {Watanabe}, {Williams}, {Wolfs}, {Wright}, {Yamada}, {Yamaguchi}, {Yamaoka}, {Yamasaki}, {Yamauchi}, {Yamauchi}, {Yanagase}, {Yaqoob}, {Yasuda}, {Yoshioka}, {Zabala}, \& {Irina}}]{xrism}
{Tashiro}, M., {Maejima}, H., {Toda}, K., {et~al.} 2018, in Society of Photo-Optical Instrumentation Engineers (SPIE) Conference Series, Vol. 10699, Space Telescopes and Instrumentation 2018: Ultraviolet to Gamma Ray, ed. J.-W.~A. {den Herder}, S.~{Nikzad}, \& K.~{Nakazawa}, 1069922, \dodoi{10.1117/12.2309455}

\bibitem[{{Titarchuk} {et~al.}(2014){Titarchuk}, {Seifina}, \& {Shrader}}]{titarchuk14}
{Titarchuk}, L., {Seifina}, E., \& {Shrader}, C. 2014, \apj, 789, 98, \dodoi{10.1088/0004-637X/789/2/98}

\bibitem[{{Tomsick} {et~al.}(2018){Tomsick}, {Parker}, {Garc{\'\i}a}, {Yamaoka}, {Barret}, {Chiu}, {Clavel}, {Fabian}, {F{\"u}rst}, {Gandhi}, {Grinberg}, {Miller}, {Pottschmidt}, \& {Walton}}]{tomsick18}
{Tomsick}, J.~A., {Parker}, M.~L., {Garc{\'\i}a}, J.~A., {et~al.} 2018, \apj, 855, 3, \dodoi{10.3847/1538-4357/aaaab1}

\bibitem[{{Ueda} {et~al.}(2005){Ueda}, {Mitsuda}, {Murakami}, \& {Matsushita}}]{ueda05}
{Ueda}, Y., {Mitsuda}, K., {Murakami}, H., \& {Matsushita}, K. 2005, \apj, 620, 274, \dodoi{10.1086/426933}

\bibitem[{{Ursini} {et~al.}(2024){Ursini}, {Gnarini}, {Capitanio}, {Bobrikova}, {Cocchi}, {Di Marco}, {Fabiani}, {Farinelli}, {La Monaca}, {Rankin}, {Saade}, \& {Poutanen}}]{ursini24}
{Ursini}, F., {Gnarini}, A., {Capitanio}, F., {et~al.} 2024, Galaxies, 12, 43, \dodoi{10.3390/galaxies12040043}

\bibitem[{{van der Klis}(2005)}]{vanderklis05}
{van der Klis}, M. 2005, in NATO Advanced Study Institute (ASI) Series B, Vol. 210, The Electromagnetic Spectrum of Neutron Stars, 283

\bibitem[{{Verner} {et~al.}(1996){Verner}, {Ferland}, {Korista}, \& {Yakovlev}}]{verner96}
{Verner}, D.~A., {Ferland}, G.~J., {Korista}, K.~T., \& {Yakovlev}, D.~G. 1996, \apj, 465, 487, \dodoi{10.1086/177435}

\bibitem[{{Vrtilek} {et~al.}(1986){Vrtilek}, {Kahn}, {Grindlay}, {Helfand}, \& {Seward}}]{vrtilek86}
{Vrtilek}, S.~D., {Kahn}, S.~M., {Grindlay}, J.~E., {Helfand}, D.~J., \& {Seward}, F.~D. 1986, \apj, 307, 698, \dodoi{10.1086/164455}

\bibitem[{{Weisskopf} {et~al.}(2022){Weisskopf}, {Soffitta}, {Baldini}, {Ramsey}, {O'Dell}, {Romani}, {Matt}, {Deininger}, {Baumgartner}, {Bellazzini}, {Costa}, {Kolodziejczak}, {Latronico}, {Marshall}, {Muleri}, {Bongiorno}, {Tennant}, {Bucciantini}, {Dovciak}, {Marin}, {Marscher}, {Poutanen}, {Slane}, {Turolla}, {Kalinowski}, {Di Marco}, {Fabiani}, {Minuti}, {La Monaca}, {Pinchera}, {Rankin}, {Sgro'}, {Trois}, {Xie}, {Alexander}, {Allen}, {Amici}, {Andersen}, {Antonelli}, {Antoniak}, {Attin{\`a}}, {Barbanera}, {Bachetti}, {Baggett}, {Bladt}, {Brez}, {Bonino}, {Boree}, {Borotto}, {Breeding}, {Brienza}, {Bygott}, {Caporale}, {Cardelli}, {Carpentiero}, {Castellano}, {Castronuovo}, {Cavalli}, {Cavazzuti}, {Ceccanti}, {Centrone}, {Citraro}, {D'Amico}, {D'Alba}, {Di Gesu}, {Del Monte}, {Dietz}, {Di Lalla}, {Persio}, {Dolan}, {Donnarumma}, {Evangelista}, {Ferrant}, {Ferrazzoli}, {Ferrie}, {Footdale}, {Forsyth}, {Foster}, {Garelick}, {Gunji}, {Gurnee}, {Head}, {Hibbard}, {Johnson}, {Kelly}, {Kilaru}, {Lefevre},
  {Roy}, {Loffredo}, {Lorenzi}, {Lucchesi}, {Maddox}, {Magazzu}, {Maldera}, {Manfreda}, {Mangraviti}, {Marengo}, {Marrocchesi}, {Massaro}, {Mauger}, {McCracken}, {McEachen}, {Mize}, {Mereu}, {Mitchell}, {Mitsuishi}, {Morbidini}, {Mosti}, {Nasimi}, {Negri}, {Negro}, {Nguyen}, {Nitschke}, {Nuti}, {Onizuka}, {Oppedisano}, {Orsini}, {Osborne}, {Pacheco}, {Paggi}, {Painter}, {Pavelitz}, {Pentz}, {Piazzolla}, {Perri}, {Pesce-Rollins}, {Peterson}, {Pilia}, {Profeti}, {Puccetti}, {Ranganathan}, {Ratheesh}, {Reedy}, {Root}, {Rubini}, {Ruswick}, {Sanchez}, {Sarra}, {Santoli}, {Scalise}, {Sciortino}, {Schroeder}, {Seek}, {Sosdian}, {Spandre}, {Speegle}, {Tamagawa}, {Tardiola}, {Tobia}, {Thomas}, {Valerie}, {Vimercati}, {Walden}, {Weddendorf}, {Wedmore}, {Welch}, {Zanetti}, \& {Zanetti}}]{ixpe}
{Weisskopf}, M.~C., {Soffitta}, P., {Baldini}, L., {et~al.} 2022, Journal of Astronomical Telescopes, Instruments, and Systems, 8, 026002, \dodoi{10.1117/1.JATIS.8.2.026002}

\bibitem[{{White} {et~al.}(1988){White}, {Stella}, \& {Parmar}}]{white88}
{White}, N.~E., {Stella}, L., \& {Parmar}, A.~N. 1988, \apj, 324, 363, \dodoi{10.1086/165901}

\bibitem[{{Wilms} {et~al.}(2000){Wilms}, {Allen}, \& {McCray}}]{wilms00}
{Wilms}, J., {Allen}, A., \& {McCray}, R. 2000, \apj, 542, 914, \dodoi{10.1086/317016}

\bibitem[{{XRISM Collaboration} {et~al.}(2024){XRISM Collaboration}, {Audard}, {Awaki}, {Ballhausen}, {Bamba}, {Behar}, {Boissay-Malaquin}, {Brenneman}, {Brown}, {Corrales}, {Costantini}, {Cumbee}, {Diaz Trigo}, {Done}, {Dotani}, {Ebisawa}, {Eckart}, {Eckert}, {Enoto}, {Eguchi}, {Ezoe}, {Foster}, {Fujimoto}, {Fujita}, {Fukazawa}, {Fukushima}, {Furuzawa}, {Gallo}, {Garc{\'\i}a}, {Gu}, {Guainazzi}, {Hagino}, {Hamaguchi}, {Hatsukade}, {Hayashi}, {Hayashi}, {Hell}, {Hodges-Kluck}, {Hornschemeier}, {Ichinohe}, {Ishida}, {Ishikawa}, {Ishisaki}, {Kaastra}, {Kallman}, {Kara}, {Katsuda}, {Kanemaru}, {Kelley}, {Kilbourne}, {Kitamoto}, {Kobayashi}, {Kohmura}, {Kubota}, {Leutenegger}, {Loewenstein}, {Maeda}, {Markevitch}, {Matsumoto}, {Matsushita}, {McCammon}, {McNamara}, {Mernier}, {Miller}, {Miller}, {Mitsuishi}, {Mizumoto}, {Mizuno}, {Mori}, {Mukai}, {Murakami}, {Mushotzky}, {Nakajima}, {Nakazawa}, {Ness}, {Nobukawa}, {Nobukawa}, {Noda}, {Odaka}, {Ogawa}, {Ogorzalek}, {Okajima}, {Ota}, {Paltani}, {Petre}, {Plucinsky},
  {Porter}, {Pottschmidt}, {Sato}, {Sato}, {Sawada}, {Seta}, {Shidatsu}, {Simionescu}, {Smith}, {Suzuki}, {Szymkowiak}, {Takahashi}, {Takeo}, {Tamagawa}, {Tamura}, {Tanaka}, {Tanimoto}, {Tashiro}, {Terada}, {Terashima}, {Tsuboi}, {Tsujimoto}, {Tsunemi}, {Tsuru}, {Uchida}, {Uchida}, {Uchida}, {Uchiyama}, {Ueda}, {Uno}, {Vink}, {Watanabe}, {Williams}, {Yamada}, {Yamada}, {Yamaguchi}, {Yamaoka}, {Yamasaki}, {Yamauchi}, {Yamauchi}, {Yaqoob}, {Yoneyama}, {Yoshida}, {Yukita}, {Zhuravleva}, {Xiang}, {Minezaki}, {Buhariwalla}, {Gerolymatou}, \& {Hagen}}]{NGC4151}
{XRISM Collaboration}, {Audard}, M., {Awaki}, H., {et~al.} 2024, \apjl, 973, L25, \dodoi{10.3847/2041-8213/ad7397}

\bibitem[{{Zdziarski} {et~al.}(2020){Zdziarski}, {Szanecki}, {Poutanen}, {Gierli{\'n}ski}, \& {Biernacki}}]{thcomp}
{Zdziarski}, A.~A., {Szanecki}, M., {Poutanen}, J., {Gierli{\'n}ski}, M., \& {Biernacki}, P. 2020, \mnras, 492, 5234, \dodoi{10.1093/mnras/staa159}

\bibitem[{{Zimmerman} {et~al.}(2005){Zimmerman}, {Narayan}, {McClintock}, \& {Miller}}]{zimmerman05}
{Zimmerman}, E.~R., {Narayan}, R., {McClintock}, J.~E., \& {Miller}, J.~M. 2005, \apj, 618, 832, \dodoi{10.1086/426071}

\end{thebibliography}

\end{document}